\newcommand{\beq}{  \begin{eqnarray}}
\newcommand{\eeq}{  \end{eqnarray}}

\documentclass[pra,twocolumn,shopacs,preprintnumbers,asmath,amssymb]{revtex4-1}

\usepackage{amsmath,amssymb}
\usepackage{subfig}
\usepackage{graphicx}
\usepackage{float}
\usepackage{dcolumn}
\usepackage{bm}
\usepackage{bbold}
\begin{document}

\title{Abelian and non-Abelian topological behavior of a neutral spin-1/2 particle in  
a  background magnetic field. }
\email{bernard@physics.unlv.edu}

\author{B. Zygelman}

\affiliation{Department of Physics and Astronomy, University of Nevada, Las Vegas, USA}

\begin{abstract}
We present results of a numerical experiment in which a neutral spin-1/2 particle subjected to a static magnetic vortex field
passes through a double-slit barrier. We demonstrate that the resulting interference pattern on a detection screen exhibits fringes reminiscent of Aharonov-Bohm scattering by a magnetic flux tube.  
To gain better understanding of the observed behavior, we provide analytic solutions for a neutral spin-1/2 rigid planar rotor in the aforementioned magnetic field. We demonstrate how that system  exhibits a non-Abelian Aharonov-Bohm effect
due to the emergence of an effective Wu-Yang (WY)flux tube. We study the behavior of the gauge invariant partition function and demonstrate a topological phase transition for the spin-1/2 planar rotor. We provide an
expression for the partition function in which
its dependence on the Wilson loop integral of the WY gauge potential is explicit. We generalize to a spin-1 system
in order to explore the Wilzcek-Zee (WZ) mechanism in a full quantum setting. We show how degeneracy can be lifted by higher order gauge
corrections that alter the semi-classical, non-Abelian, WZ phase. 
Models that allow analytic description offer a foil to objections that question the fidelity of
predictions based on the generalized Born-Oppenheimer approximation in atomic and molecular systems.

Though the primary focus of this study concerns the emergence of gauge structure in neutral systems, the theory is also applicable to systems that posses electric charge. In that case, we explore interference between fundamental gauge fields (i.e. electromagnetism) with effective gauge potentials. We propose a possible laboratory
demonstration for the latter in an ion trap setting. We illustrate
how effective gauge potentials influence wave-packet revivals in the said ion trap.

\end{abstract}

\maketitle
\section{Introduction}
The double slit experiment and the Aharonov-Bohm (AB) effect\cite{ab59} are iconic examples that highlight novel and counter-intuitive aspects of the 
quantum theory\cite{ball18}.
 The former has long served as a pedagogical device\cite{feynman2011} to introduce the notion of wave-particle duality to students of quantum mechanics and laboratory demonstrations of it have raised new questions regarding the role of measurement in quantum mechanics (QM) {\cite{Peruzzo634,delayed2000}. The AB effect demonstrates the role of gauge potentials in quantum mechanics, and Feynman\cite{feynman2011} framed it in a double slit setting to illustrate and underscore its topological significance. 

From the Einstein-Bohr-Sommerfield quantization rules to the TKNN integers\cite{thoull82}, topology has always played a role in QM, and for which the AB effect offers an instructive template. It has been applied to elaborate on the nature of anyons\cite{Wilczek82} and other forms of exotic quantum matter\cite{TOPI2010}. Researchers hope to harness topology in service of enabling high-fidelity qubit technology\cite{sarma06} and fault tolerant quantum computing\cite{preskill1997}. 

In this paper we illustrate how AB-like topological effects, and its non-Abelian generalization\cite{wuyang75,horvath86}, manifest in simple quantum systems that allow accurate numerical as well as analytic solutions. First, we consider the dynamics of a neutral spin-1/2 system coupled to an external static magnetic field. We
perform a quantum mechanical numerical experiment in which the particle passes through a double-slit barrier. When the position of the particle is measured at a detection screen we find an anticipated wave interference pattern. 

In addition to interference due to the presence of slit barriers, we show that the resulting pattern is best described by  appealing to a model in which a charged particle is minimally coupled
to an effective magnetic flux tube. This, despite the fact that the spin-1/2 particle is neutral and couples locally to the external field via the standard $ {\vec  {\mu}} \cdot {\vec B}$ term.

Our numerical experiment provides a demonstration of how effective gauge potentials arise in quantum system that appear to have no overt gauge structure. This system (without the double slit) was first proposed\cite{march92} as an example of inertial frame dragging. Here we confirm, via our numerical simulation, the predictions of that gedanken system. In addition to the predicted\cite{march92} Abelian AB behavior, we explore non-Abelian features inherent in analogous systems that allow analytic solution.

In section II, we summarize the results of our numerical experiment. We demonstrate the scattering of a neutral spin-1/2 wave-packet by a double slit barrier. The packet experiences a background magnetic field ${\vec B}$ in which the condition  
 ${\vec \nabla } ({\vec \mu} \cdot {\vec B}) =0 $, is satisfied. The latter insures that the packet does not experience a gradient force. We analyze the interference pattern 
at a post-slit detection screen and find that it shares the predicted structure of a charged particle that is scattered by an AB magnetic flux tube.

In order to gain better understanding of this phenomenon, we introduce, in section III, a system that allows analytic solution. We calculate the partition function of a neutral spin-1/2  planar rotor placed in the aforementioned $\vec B$ field configuration. In addition to verifying the AB features observed in our numerical demonstration, we conclude that a model characterized by a non-Abelian Wu-Yang\cite{wuyang75} (WY) flux tube provides a more accurate description. We demonstrate that the, gauge invariant, partition function is an explicit function of the Wilson-loop\cite{Mak09} integral of a (WY) gauge field. 

Early studies\cite{mead76,moo86,zyg87a,jackiw88} have demonstrated how non-trivial gauge structures arise in molecular and atomic systems. In  low energy atomic collisions\cite{zyg87a,zyg90} and molecular structure\cite{moo86} calculations, it is convenient to express the state vector in a basis
of Born-Oppenheimer eigen-states. A complete set of such states leads to gauge potentials, coupled to the nuclear motion, that have both spatial and temporal components\cite{zyg87a,zyg90,zyg15}. The spatial components describe a pure gauge, and its is only after truncation from a Hilbert space spanned by a complete set to a subspace that the spatial components acquire a non-trivial Wilson-loop value. For that reason it has sometimes been argued that gauge fields that lead to non-trivial Wilson loop integrals, (a.k.a geometric, or Berry, phases) are artifacts of the approximation or truncation procedure. In section IV, we investigate this question for the model introduced in section II. We demonstrate how an open ended, but gauge invariant, Wilson-line integral of a $3+1$ gauge field along a space-time path can lead to a non-trivial spatial Wilson loop integral when projected to a closed path of the spatial subspace. 

 Wilzcek and Zee\cite{zee} demonstrated how non-Abelian geometric phases arise in the slow evolution of a system possessing degenerate adiabatic eigen-states that are well separated from distant states. 
As our spin-1/2 model contains only two internal states, separated by an energy gap, the Wilczek-Zee mechanism is not applicable. Therefore we introduce, in section V, an extension to our two state model by positing a three-internal
 state system that allows analytic solutions. In the latter, two internal states are degenerate and a third  state is separated from them by a large energy defect. We analyze its gauge structure, and show that higher order gauge corrections\cite{zyg86,zyg87a,ber89,zyg90} breaks the degeneracy evident in (semi-classical) adiabatic evolution\cite{zee}. As a consequence, gauge covariance is regained only in the 3+1 formalism\cite{zyg87a}. In section VI, we provide a summary and conclusion of our efforts and propose possible systems in which the effects described above may be gleaned in a laboratory setting.
 
 Unless otherwise stated we use units in which $\hbar=1$. With the exception of the Pauli matrices, we use boldface typeface to represent both vector and matrix valued quantities. In some cases, when there is the possibility of ambiguity, we use explicit vector notation to represent vector valued quantities. 

\section{Numerical double-slit experiment for a neutral spin-1/2 system in a static magnetic field}
Consider a neutral spin 1/2 atom or neutron with magnetic moment $ {\bm \mu}$, and mass $m$,
in the presence of a static background magnetic field 

\beq
 {\vec B}= B(\rho) \, {\bm {\hat \phi}} + B_{0} \, {\bm {\hat k} }
\label{0.1}
\eeq
where $\phi,\rho$ are the polar  and radial coordinates in a cylindrical coordinate system. 
We take $ B(\rho) \equiv B_{\rho}$, and $ B_{0}$ to be constants so that ${\vec B}$ describes a vortex configuration
superimposed on a constant magnetic field in the ${\bm \hat{k}}$ direction. 
The Hamiltonian for  a neutral spin-1/2 system is 
\beq
H= - \, \frac{\hbar^{2}}{2m} \mathbb{1} \, {\vec \nabla}^{2}_{\vec R} + \mu \,  {\vec {\sigma}} \cdot {\vec B} \label{0.2}
\eeq
where $ \mathbb{1}$ is the unit $2 \times 2$ matrix and $ {\vec {\sigma} }$ are Pauli matrices.
The adiabatic, or BO, eigenergies of $H$ are the constant surfaces
\beq 
V_{BO}= \pm \mu \sqrt{B_{\rho}^{2}+B_{0}^{2}} \equiv \pm \Delta
\label{0.3}
\eeq
separated by a finite energy gap $2\Delta$.
Though the magnetic
field lines have a vortex structure, and ignoring a small higher order correction\cite{zyg15}, 
the gradient force $ - {\vec\nabla} V_{BO} $ vanishes. Thus wave packets evolve, as confirmed in a previous numerical study\cite{zyg15}, 
with minimal distortion induced by the presence of scalar potentials.

Fig. (\ref{fig:fig1}) describes a wave packet, initially in the ground adiabatic state, 
whose probability density, as a function of time, is illustrated in the panels of that figure.
In the first run of a simulation we set $ B_{\rho}=0$ and the system evolves on the ground state adiabatic
surface as the particle proceeds through the two slits.
At the detection screen, shown by the red dashed line, 
the wave amplitude forms an interference  pattern whose probability density is plotted in the left panel of Fig. (\ref{fig:fig2}). In that figure the solid blue line represents the data of this numerical simulation
whereas the red dashed line is an analytic fit to the simulation.
In calculating the latter we assumed that the probability amplitude at the observation screen is given by
\beq
\psi = \psi_{R} + \exp(i \beta)\psi_{L}
\label{0.4}
\eeq
where $\psi_{R,L}$ are amplitudes, based on a Huygens principle construction, due to contributions coming from the right and left slits, shown in Fig. (\ref{fig:fig3}), respectively. $\beta$ is a measure of the relative phase between
the amplitudes and for this run
$\beta \approx 0$ provides the best fit.
On a second run we set $ B_{0} =0,  B_{\rho} = |\Delta| $  so that the Zeeman
energy splittings are unchanged from that of the first run.  The resulting interference pattern is illustrated on the second (r.h.s) panel of Fig.(\ref{fig:fig2}) by the red line, and in that case we found the best value for $\beta\approx \pi$. 
In a subsequent run
we translated the $ {\vec B}$ field so that the vortex center, labeld $x_{c} $ on the horizontal axis of Fig. (\ref{fig:fig3}),  has been shifted to a point that
is not framed by the pair of slits in the barrier. In that simulation we again
found that $\beta \approx 0$ provides the best fit to the numerical data.
We also considered different ratios  $\tan\theta = B_{\rho}/B_{0}$ and fit $ \beta $ for  these  choices
of $\theta$. The results are summarized by the following observations,
\begin{enumerate}
\item{ The data obtained in the simulations, for vortex centers $ -L/2 < x_{c} < L/2$,  are best described by Eq. (\ref{0.4}) provided that
$ \beta $ takes the value $ \pi (1- \cos \theta)$.}
\item{For an external magnetic field in which $ |x_{c}| > |L/2| $ the value $ \beta \approx 0 $ provides the best fit.}
\item{If the packet mean kinetic energy  $E >> 2 \Delta$ the interference pattern is largely insensitive to 
the location of $x_{c}$ and is best fit with $\beta \approx 0$. }
\end{enumerate}
The features described above are suggestive of dynamics influenced by topology. Indeed, it is the behavior predicted in Feynman's 
thought experiment treatment of Aharonov-Bohm (AB) scattering\cite{ab59} of a charged scalar particle in a double slit apparatus\cite{feynman2011}. Observations (1-3)
are consistent with the following hypothesis,
\beq
  \beta = \oint_{C}  d{\vec s} \cdot {\vec A}   \quad {\rm \, where } \quad  {\vec A} =  
\frac{(1-\cos\theta ) }{2 \rho} \,  {\bm {\hat \phi}}
\label{0.5}
\eeq
is a gauge potential that describes Aharonov-Bohm (AB)-like flux tube of strength $ (1-\cos\theta )/2$ centered on the barrier at
$x_{c}=0$.
The line integral is taken along a single circuit about a closed path $C$ that circumscribes $x_{c}=0$ on the barrier.

Hamiltonian (\ref{0.2}) possesses no overt gauge structure, but it is known \cite{zyg87a,jackiw88,moo86,mead76} that 
effective gauge potentials can emerge in quantum systems not coupled to fundamental gauge fields. In this study we highlight the utility of using a gauge theory framework to
characterize quantum systems that exhibit apparent topological AB-like behavior in a scattering setting. However, the features itemized above do not completely fit into the standard AB framework. It requires, as shown below, application of non-Abelian ideas and in order
to elaborate on this observation we introduce a simpler physical system that allows an analytic
description.

\begin{figure}[ht]
\centering
\includegraphics[width=0.9 \linewidth]{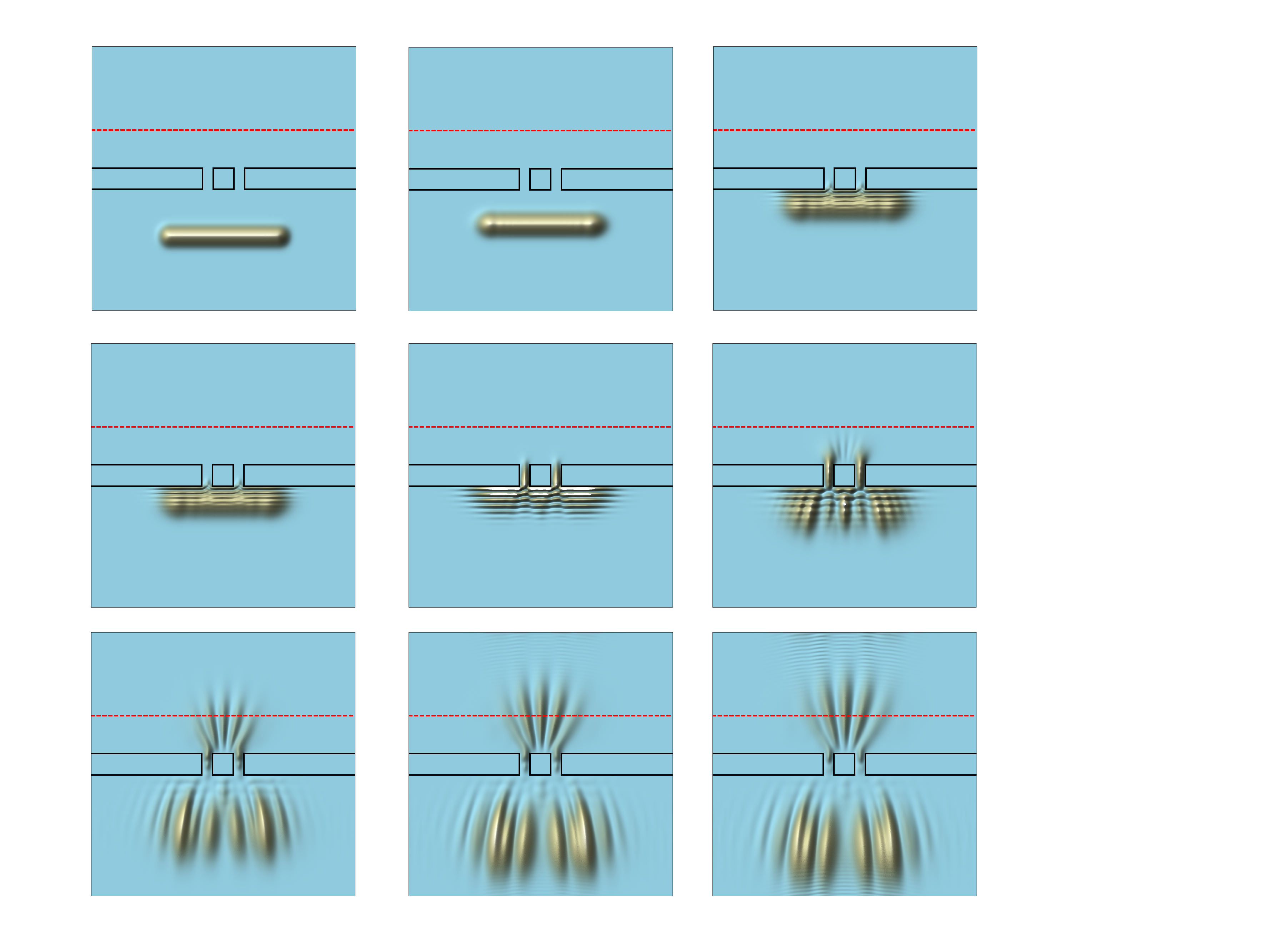}
\caption{\label{fig:fig1} Time series plot of wave packet at initial time $t_{0}$ as it proceeds to, from left to right,
to a double slit barrier (solid line outline). The red dashed line represent a detection screen, and at final time $t_{f}$ 
shown in the panel at the lower right, the particle
position is measured.}
\end{figure}

\begin{figure}[ht]
\centering
\includegraphics[width=0.7 \linewidth]{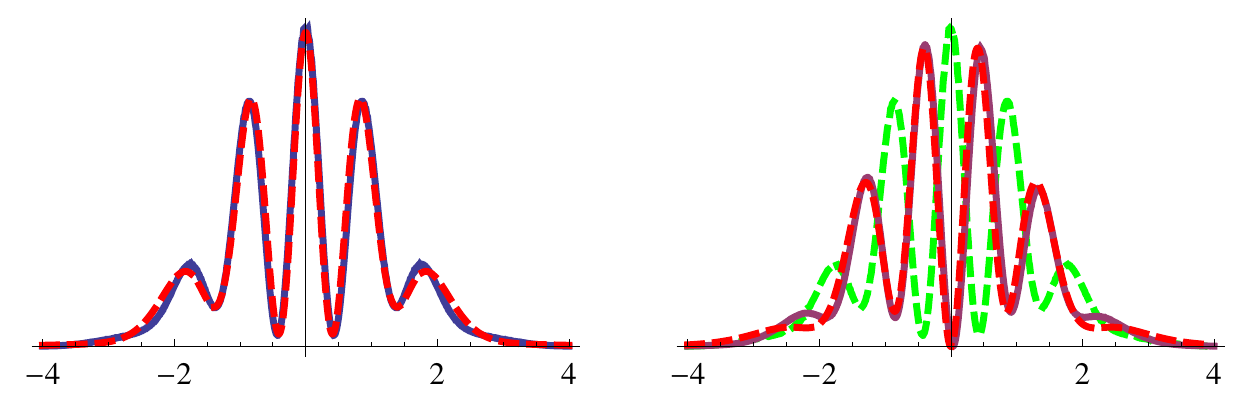}
\caption{\label{fig:fig2} Comparison of observed interference patterns at observation screen with a fit to model Eq. (\ref{0.4}).
Blue lines represent simulation data, red dashed lines represents fit to Eq. (\ref{0.4}).
The panel on the left corresponds to case where $B_{\rho}=0$ or $\theta=0$. 
The panel on the right corresponds to case where $\theta=\pi/2 \, (B_{0}=0)$, and the green line represents fringe patterns corresponding to $\beta=0$.}
\end{figure}

\begin{figure}[ht]
\centering
\includegraphics[width=0.8 \linewidth]{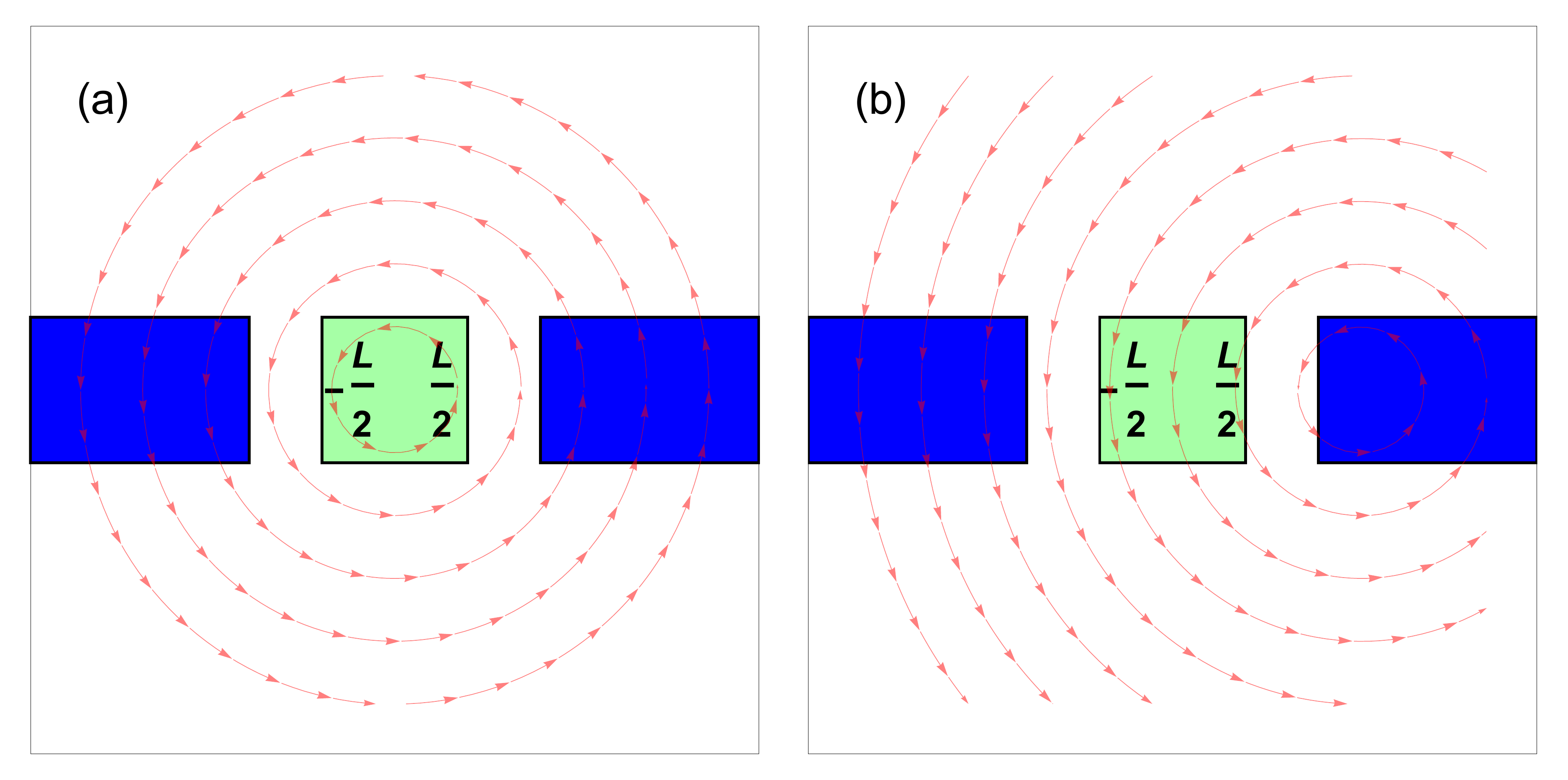}
\caption{\label{fig:fig3} 
Translation of magnetic field vortex center. Panel (a) vortex center $x_{c}$ is situated at the center of the double-slit configuration.
In Panel (b) $x_{c}$ is translated to the right. }
\end{figure}

\section{The spin-1/2 rotor; an analytic treatment.}
We substitute the 2D kinetic energy operator, in Eq. (\ref{0.2}),  
$ \frac{\hbar^{2}}{2m} \, {\vec \nabla}^{2}_{\vec R} 
\rightarrow   \frac{1}{2 I} \, \partial_{\phi}^{2} $ (setting $\hbar=1$)  so that
\beq
H = -\frac{1}{2 I} \,\mathbb{1} \, \partial_{\phi}^{2} \,  +  \mu \, \sigma \cdot {\vec B} .
\label{1.01a}
\eeq
$H$  describes a neutral spin-1/2 particle constrained on a unit circle, ( i.e. a free rotor with spin and moment of inertia $I$), subjected to an external magnetic given in Eq. (\ref{0.1}). The rotor coordinates $\phi=0,2\pi$ are identified.
Hamiltonian Eq. (\ref{1.01a}) can be re-written as
\beq
&& {\bm H} = -\frac{1}{2 I} \partial^{2}_{\phi}  + {\bm V }  \nonumber \\
&&  {\bm V}  =  \left(
\begin{array}{cc}
\Delta \cos\theta & - i \exp(-i \phi ) \, \Delta \sin\theta \\
 i \exp(i \phi) \, \Delta \sin\theta & - \Delta \cos \theta 
\end{array}
\right).  \label{1.02}
\eeq
${\bm H}$ commutes with 
\beq 
{\bm J} = -i \frac{\partial}{\partial \phi} + \frac{1}{2} \, \sigma_{3}  
\label{1.03}
\eeq
whose eigenstates  
\beq
\psi = \left(
\begin{array}{c}
\exp(i (m-1) \phi) \, c_{1} \\
 \exp(i m \phi)\,  c_{2} 
\end{array}
\right),  
\label{1.04}
\eeq
where $m$ is an integer, satisfy
$$ {\bm J} \psi = (m-1/2)  \psi. $$
Using ansatz (\ref{1.04})
we find that the eigenvalue equation $ ({\bm H} -E) \psi=0$ reduces to
$$ ({\cal H}-E ) {\underbar c} =0 $$ 
where $ {\underbar c} \equiv \left(
\begin{array}{c}
 c_{1} \\
  c_{2} 
\end{array} \right )$
and
\beq
&& {\cal H} =  \left ( \frac{m^2}{2 I} +\frac{1-2m}{4 I} \right ) \mathbb{1}
+ \nonumber \\
&& \left ( \frac{1-2m}{4 I} + \Delta \cos\theta \right ) \sigma_{3}
+ \Delta \sin\theta \, \sigma_{2}. 
\label{1.05}
\eeq
$  \mathbb{1} $ is the unit matrix and $ \sigma_{2}, \sigma_{3} $ are
Pauli matrices.
Introducing the unitary operator $ W = \exp(-i \sigma_{1} \Omega/2) $ 
where
\beq
&& \cos\Omega = \frac{ \frac{(1-2 m)}{4 I}  + \Delta \cos\theta} 
{\sqrt{\frac{(1-2 m)^2}{16 I^2} + \frac{(1-2 m) \Delta \cos\theta}{2 I} + \Delta^{2}} } \nonumber \\
&& 
\sin \Omega = \frac{\Delta \sin\theta}
{\sqrt{\frac{(1-2 m)^2}{16 I^2} + \frac{(1-2 m) \Delta \cos\theta}{2 I} + \Delta^{2}} },
\label{1.06}
\eeq
we find that
\beq
&& W {\cal H} W^{\dag} = 
\left ( \frac{m^2}{2 I} +\frac{1-2m}{4 I} \right ) \mathbb{1} + \nonumber \\
&& {\sqrt{\Delta^{2}+\frac{(1-2 m)^2}{16 I^2} + \frac{(1-2 m) \Delta \cos\theta}{2 I}} } \, \sigma_{3}
\label{1.07}
\eeq
Therefore,
\beq
\psi_{\pm} = 
\left (  \begin{array}{cc}
 \exp(i (m-1) \phi)  & 0\\
 0  & \exp(i m \phi) 
\end{array} \right )  \, W^{\dagger}  \, 
| \pm \rangle, 
\label{1.11}
\eeq
where
$$
| + \rangle = \left ( \begin{array}{c}
 1 \\
 0 
\end{array} \right )
\quad
| -\rangle = \left ( \begin{array}{c}
 0 \\
 1 
\end{array} \right ),
$$
are eigenstates of $ {\bm H} $.
That is, 
$$ {\bm H} \psi_{\pm}(m) = E_{\pm}(m) \psi_{\pm}(m) $$
where
\beq
&& E_{\pm} = \frac{m^{2}}{2 I} + \frac{(1-2 m)}{4 I} \pm \nonumber \\
&& {\sqrt{\frac{(1-2 m)^2}{16 I^2} + \frac{(1-2 m) \Delta \cos\theta}{2 I} + \Delta^{2}} }
\label{1.12}
\eeq
In the limit $ \Delta \rightarrow 0 $, and for $ 1 - 2 m > 0 $, $ \cos\Omega \rightarrow 1 $,
\beq
&& \psi_{+} \rightarrow \exp(i (m-1) \phi) |+ \rangle  \quad E_{+} \rightarrow \frac{(m-1)^2}{2 I} \nonumber \\
&& \psi_{-} \rightarrow \exp(i m \phi) |- \rangle  \quad E_{-} \rightarrow \frac{m^2}{2 I}, 
\label{1.13}
\eeq
likewise, for $ 1 - 2 m < 0 $ $ \cos\Omega \rightarrow -1 $, and
\beq
&& \psi_{+} \rightarrow \exp(i m \phi) |+ \rangle  \quad E_{+} \rightarrow \frac{m^2}{2 I} \nonumber \\
&& \psi_{-} \rightarrow \exp(i (m-1) \phi) |- \rangle  \quad E_{-} \rightarrow \frac{(m-1)^2}{2 I}.
\label{1.14} 
\eeq
Eqs. (\ref{1.13},\ref{1.14}) correspond to free rotor solutions.

In the limit $ \Delta \rightarrow \infty $, provided that $ \Delta > \frac{|1-2 m|}{4I} $, 
\beq
&&  E_{+}(m) \approx  m^{2} \left (\frac{1}{2 I} + \frac{\sin^2\theta}{8 I^2 \Delta} \right )
- m \left ( \frac{( 1+\cos\theta) }{2 I} + \frac{\sin^2 \theta}{8 I^2 \Delta} \right ) + \nonumber \\
&& \Delta + \frac{1+\cos\theta}{4I} + \frac{\sin\theta^{2}}{32 I^2 \Delta} + {\cal O}(\frac{1}{\Delta^{2}}) \,  \dots
\label{1.15}
\eeq
and
\beq
&& E_{-}(m) \approx  m^{2} \left ( \frac{1}{2 I} - \frac{\sin^2\theta}{8 I^2 \Delta} \right )
- m \left ( \frac{( 1- \cos\theta) }{2 I} - \frac{\sin^2\theta}{8 I^2 \Delta} \right ) - \nonumber \\
&&\Delta + \frac{1-\cos\theta}{4I} - \frac{\sin\theta^{2}}{32 I^2 \Delta} + {\cal O}(\frac{1}{\Delta^{2}}) \,  \dots
\label{1.16}
\eeq

\subsection{Adiabatic gauge}
In order to gain insight into these solutions we transform the eigenvalue equation corresponding to
Hamiltonian (\ref{1.02}) into the so-called adiabatic representation \cite{zyg87a} which we define by
\beq
\psi = U \, F 
\label{3.00}
\eeq
where 
\beq
U \equiv \exp(-i \sigma_{3} \phi/2) \exp(i \sigma_{1} \theta/2) \exp(i \sigma_{3} \phi/2)
\label{3.01}
\eeq
is a single-valued unitary operator.
We get
\beq
-\frac{1}{2I} ( \partial_{\phi} - i {\bm A})^{2} F + \Delta \, \sigma_{3} F = E F
\label{3.02}
\eeq
where the non-Abelian, pure, gauge potential
\beq
&& {\bm A} = i U^{\dag} \partial_{\phi}  U =  \nonumber \\
&&  \frac{1}{2} \left ( \begin{array}{cc}   \cos\theta -1 & i \, \sin\theta \exp(-i \phi) \\
-i \, \sin\theta \exp(i \phi) & (1- \cos\theta) \end{array} \right ).
\label{3.03}
\eeq
If we ignore the off-diagonal components of the gauge potential and project this equation to the ground 
manifold via projection operator $ |- \rangle \langle - |$,  we find
\beq
&&  -\frac{1}{2I}
( \partial_{\phi} - i A_{g} )^{2} F_{g} + \frac{\beta}{2I} F_{g} - \Delta F_{g} = E F_{g}  \nonumber \\
&&   A_{g} = 1/2 (1- \cos\theta) \equiv \alpha\nonumber \\ 
&&   \beta = A_{12} A_{21} = \sin^{2}\theta/4 = \alpha(1-\alpha). 
\label{3.04}
\eeq
We note that
\beq
F_{g} = \exp(i m \phi)  
\label{3.05}
\eeq
is an eigenstate of Eq. (\ref{3.04}) corresponding to eigenvalue
\beq
&& E_{g} = \frac{(m-\alpha)^{2}}{2 I} + \frac{(\alpha -\alpha^{2} )}{2 I} - \Delta = \nonumber \\
&&  \frac{m^{2}}{2 I} - \frac{m \, \alpha}{ I} + \frac{\alpha}{2 I} - \Delta. = \nonumber \\
\label{3.08}
\eeq
It agrees with the leading order limit of expression (\ref{1.16}) as $ \Delta \rightarrow \infty$,
\beq
E_{-}= \frac{m^{2}}{2 I} -\frac{ m \, \alpha}{I} + \frac{\alpha}{2 I} - \Delta = E_{g},
\label{3.09}
\eeq

Consider now the excited state manifold obtained via projection $ |+ \rangle \langle +| $. 
\beq
F_{e} = \exp(i (m-1) \phi) 
\label{3.10}
\eeq
is an eigenstate of the latter corresponding to eigenvalue
\beq
&& E_{e} = (m-1 + \alpha)^{2}+ \frac{\beta^{2}}{4 I} + V = \nonumber \\
&& E_{e} = \frac{m^{2}}{2 I} + \frac{ m (\alpha-1)}{I} +\frac{1-\alpha}{2 I}
\label{3.11}
\eeq
Note that
$$ 1-\alpha = \frac{1}{2}(1+\cos\theta) $$
and so
\beq
E_{e} =    \frac{m^{2}}{2 I} - \frac{m (1+\cos\theta)}{2 I} + \frac{1+\cos\theta}{4 I} + \Delta
\label{3.12}
\eeq
Or comparing to Eq. (\ref{1.15})  we find, as $ \Delta \rightarrow \infty$,   $ E_{e} = E_{+}$.

In conclusion, we find that in the adiabatic gauge the following solutions to Eqs. (\ref{3.02})
disregarding
the off-diagonal couplings predicts adiabatic gauge eigen-solutions
\beq
&& \psi^{a}_{g} = \exp(i m \phi ) \left ( \begin{array}{c} 0 \\ 1 \end{array} \right ) \nonumber \\
&& \psi^{a}_{e} = \exp(i (m-1) \phi)  \left ( \begin{array}{c} 1 \\  0 \end{array} \right )
\label{3.13}
\eeq
with eigen-energies $ E_{-}, E_{+}$, respectively. They agree with the leading order, in the limit $\Delta \rightarrow \infty$,
eigenvalues obtained given by the exact analytic solutions to Eq. (\ref{1.02}).
\section{ The Wu-Yang flux tube.}

Some time ago, T.T. Wu and C.N. Yang \cite{wuyang75} entertained the notion of a non-Abelian Aharonov-Bohm effect. They postulated a non-Abelian flux tube that may allow, if found in nature,  topological transformation of isotopic charge when
a system, described by an isotopic amplitude, is transported about the flux tube. In this paper we demonstrate how the spin-1/2 system
described in the previous section possesses some of the salient features of a particle, with spin degrees of freedom,
coupled to a Wu-Yang (WY) non-Abelian flux tube. To set the stage for that discussion we first introduce an idealized model in which
a free rotor is coupled to a WY connection. 
\subsection{Rotor coupled to  Wu-Yang gauge potential}
Consider the following non-Abelian gauge potential
\beq
&& { {\bm A} }= \sigma_{3} \, {\vec A}  \quad   {\vec A} = \{  A_{x}, A_{y} \}  \nonumber \\
&& \quad A_{x} =  \frac{\alpha \,  -y}{(x-x_{0})^2 +y^2 }  \quad A_{y} =
 \frac{ \alpha \, (x-x_{0})}{(x-x_{0})^2 +y^2 }
\label{S.08}
\eeq
\begin{figure}[ht]
\centering
\includegraphics[width=0.8 \linewidth]{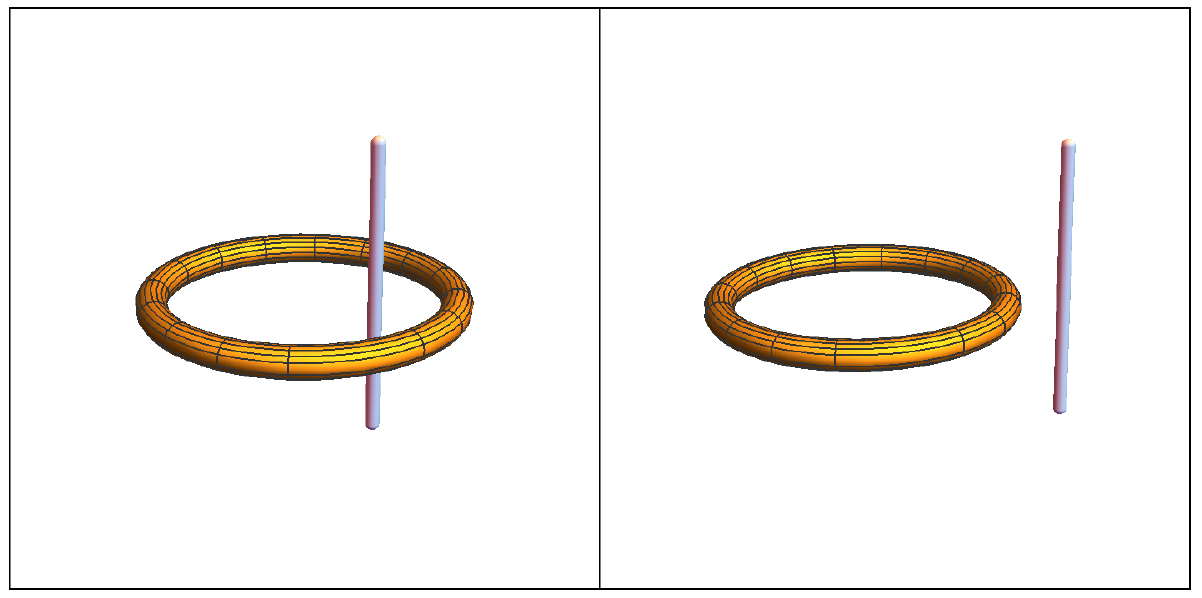}
\caption{\label{fig:fig4} Wu-Yang flux tube circumscribed by a rotor track (left panel). Right panel shows the flux tube exterior of the rotor track. }
\end{figure}
where $x,y$ are the coordinates of a (iso) spin-1/2 particle. It is straightforward to verify that the spatial
components of the matrix-valued curvature two-form ${\bm F}$ vanish identically in the region excluding the point $x=x_{0} \geq 0,y=0$. From that observations it may appear that gauge connection (\ref{S.08}) corresponds to that of a pure gauge. Nevertheless, as for the conventional AB vector potential, its Wilson loop integral circumscribing the point $x_{0},y=0$ is non-trivial.

For connection (\ref{S.08}) the gauge invariant trace of the Wilson loop phase integral has the value
\beq
W(C) \equiv Tr P \exp (-i \oint_{C}   d {\bm s} \cdot  { {\bm A} } ) = 2 \cos 2 \, m \, \pi \alpha
\label{S.09}
\eeq
where $ C$ is an arbitrary contour (of counter-clockwise sense) that encloses 
the point $(x_{0}, y=0)$ and $m$, the winding number, itemizes the number of circuits taken around $C$. $P$ represents path ordering.

As first pointed out by Wu and Yang, gauge potential (\ref{S.08}) is a non-Abelian generalization
of the Aharonov-Bohm potential.  Despite the fact that in the gauge in which
${ {\bm A}} $  is diagonal and therefore has an ``Abelianized '' structure, it is not simply the potential of two  AB flux
tubes of opposite charge\cite{comment01}. In this sense ${ {\bm A}} $  describes a non-Abelian flux tube piercing the
$x \, y$ plane at the point $ (x_{0},y=0) $

We seek a Schr{\"o}dinger equation
for a spin-1/2 particle, constrained on the unit circle $x^2+y^2=1$, coupled to gauge potential (\ref{S.08}), as well as a scalar potential $ {\bm A}_{0} = -\sigma_{3} \, \Delta $, where $\Delta $ is a constant energy defect. Constrained systems typically involve singular Lagrangians\cite{Dirac64} and a rigorous derivation of the corresponding Hamiltonian requires application of Dirac's theory\cite{Dirac64} of constrained dynamical systems. The latter has been applied to construct the quantum Hamiltonian  of a scalar particle constrained on a circular path\cite{anton02}. Here we use a more heuristic approach by considering the standard (unconstrained) Schr{\"o}dinger equation in two dimensions and in which the spin-1/2 particle is minimally coupled to gauge potential (\ref{S.08}). 
We have
\beq
-\frac{1}{2 m} ( {\bm \nabla} - i {\bm A} )^2 \psi - {\bm A}_{0} \psi
= i\, \frac{\partial \psi}{\partial t}  \label{S.10}
\eeq
where
\beq
{\vec A} = && -{\hat r} \, \frac{\alpha  x_{0} \sin (\phi )}{r^2-2 r x_{0} \cos (\phi )+{x_{0}}^2} + \nonumber \\
&& {\hat \phi} \, \frac{\alpha  (r-x_{0} \cos (\phi ))}{r^2-2 r x_{0} \cos (\phi )+{x_{0}}^2} \label{S.10a}
\eeq
is expressed in a polar coordinate system.
If $ x_{0} < r $, and in the range $ -\pi < \phi \leq \pi $, the function
\beq
\Omega_{>} = \alpha \, \frac{\phi}{2} - \alpha \, \arctan\Bigl [\frac{r+x_{0}}{r-x_{0} } \tan(\frac{\phi}{2}) \Bigr ]
\label{s10.aa}
\eeq
is single-valued and we are allowed the gauge transformation
\beq
&& \psi \rightarrow \psi' = \exp(-i \, \Omega_{>} \, \sigma_{3} ) \psi \nonumber \\
&& {\bm A} \rightarrow {\bm A}' = {\bm A} + \nabla \Omega_{>} =
{\hat \phi}\, \frac{\alpha}{r}. 
\label{s10.b}
\eeq

Thus $ {\bf A}'$ describes a WY flux tube centered
at the origin. If the particle is constrained to move on the
unit circle and $ x_{0} < 1 $  we obtain the Schr{\"o}dinger equation

\beq
\frac{1}{2 I} (\partial_{\phi} - i \,\alpha \, \sigma_{3} )^2 \psi' 
- {\bm A}_{0} \psi' = i \, \frac{\partial \psi'}{\partial t}
\label{s10.3}
\eeq

The energy eigenstates to Eq. (\ref{s10.3}) are
\beq
&& \psi'_{m}(E_{+}) =  \exp(i \, m \phi)  \left ( \begin{array}{c} \frac{1}{\sqrt{2 \pi}}  \\ 0 \end{array} \right ) \nonumber \\
&&  E_{+} = \frac{(m+\alpha)^2}{2I} + \Delta  \nonumber \\
&& \psi'_{m}(E_{-}) =  \exp(i \, m \phi) 
\left ( \begin{array}{c} 0 \\  \frac{1}{\sqrt{2 \pi}}  \end{array} \right ) 
\nonumber \\
&&  E_{-} = \frac{(m-\alpha)^2}{2I} - \Delta  
\label{S.10b}
\eeq
where $ m$ is an integer. 

For $x_{0} > r$ , $ \Omega_{>} $ is no longer single-valued but 
\beq
\Omega_{<} = -\alpha \, \frac{\phi}{2} - \alpha \, \arctan\Bigl [\frac{r+x_{0}}{r-x_{0} } \tan(\frac{\phi}{2}) \Bigr ]
\label{s10.c}
\eeq
is. Replacing $ \Omega_{>}$ with
$ \Omega_{<}$ in (\ref{s10.b})
we find $ {\bf A'} =0 $, i.e. a pure gauge. Thus,
for $ x_{0} > 1 $ Eq. (\ref{s10.3})) is replaced with
\beq
\frac{1}{2 I} { \partial^{2}}_{\phi}  \psi' 
- {\bm A}_{0} \psi' = i \, \frac{\partial \psi'}{\partial t}
\label{s10.cc}
\eeq
and,

\beq
&& \psi'_{m}(E_{+}) =  \exp(i \, m \phi)  \,  \left ( \begin{array}{c} \frac{1}{\sqrt{2 \pi}}  \\ 0 \end{array} \right ) 
\nonumber \\
&&  E_{+} = \frac{m^2}{2I} + \Delta  \nonumber \\
&& \psi'_{m}(E_{-}) =  \exp(i \, m \phi)   \left ( \begin{array}{c} 0 \\  \frac{1}{\sqrt{2 \pi}}  \end{array} \right ) 
\nonumber \\
&&  E_{-} = \frac{m^2}{2I} - \Delta.  
\label{S.11b}
\eeq
As the position of flux tube shifts from $ x_{0} < 1 $ to $ x_{0} >1 $ the  energy spectrum shifts into that of a free rotor. This topological
feature is most clearly evident in the behavior of the partition function $ {\cal Z} = \sum_{m} \, \exp(-\beta \, E_{m}) $ where $\beta$
is an inverse temperature and $E_{m}$ are the energy eigenvalues for the eigenstates summarized above.
Consider the propagator for Schr{\"o}dinger Eq. (\ref{s10.3}) in the region $ |x_{0}| < 1 $,
\beq
&& G(\phi \, t; \phi' \, t') \equiv \langle \phi | \exp(-i H \, \tau) | \phi' \rangle = \nonumber \\
&& \sum_{m} \psi'_{m}(E_{+},\phi) {{\psi'}_{m}}^{\dagger}(E_{+},\phi') \exp(-i E_{+} \tau) + \nonumber \\
&& \sum_{m} \psi'_{m}(E_{-},\phi) {{\psi'}_{m}}^{\dagger}(E_{-},\phi') \exp(-i E_{-} \tau) 
\label{R.01}
\eeq
where $ \tau = t-t'$. Thus
\beq
&&  G(\phi \, t; \phi' \, t') = \frac{1}{2 \pi}  \exp(-i \frac{\alpha^2}{2 I} \tau) \exp(-i \Delta \sigma_{3} \tau) \times \nonumber \\
 && \sum_{m} \exp(i\, m (\phi-\phi'))  \exp(-i \, \frac{m^2 \tau}{2 I}) \exp(-i \,\sigma_{3} \, m \, \frac{\alpha \tau}{I}). \nonumber \\
 \label{R.01b}
\eeq
With the following definition of the Jacobi-theta function \cite{bellman,shulman} 
\beq
\theta_{3}(z,u) \equiv \sum_{m} \exp(i \pi m^2 u) \exp(2\, i \, m \, z) 
\label{R.03}
\eeq
we re-express 
\beq
&& G(\phi \, t; \phi' \, t') \equiv 
\frac{1}{2 \pi}  \exp(-i \frac{\alpha^2 \, \tau}{2 I} ) \exp(-i \Delta \sigma_{3} \tau) \times \nonumber \\
&&  
\left ( \begin{array}{cc} 
\theta_{3}(z_{-},u)  & 0 \\
0 & \theta_{3}(z_{+},u)  \end{array} \right ) 
\label{R.04}
\eeq
where
\beq
 z_{\mp} = (\phi - \phi')/2 \mp \frac{ \, \alpha \, \tau}{2 I} 
 \quad u = - \frac{\tau }{2 \pi I}. 
\label{R.05}
\eeq
Employing the identity\cite{bellman}, 
\beq
&& \theta_{3}(z,u) = \frac{1}{\sqrt{-i \,u}} \exp(-i \frac{z^2}{\pi u}) \theta_{3}(-\frac{z}{u},-\frac{1}{u}) =\nonumber \\
&& \frac{ \exp(\frac{-i \, z^2}{\pi u})}{\sqrt{-i \, u} } \sum_{m} \exp(-\frac{i \pi \, m^2}{u} ) \exp(\frac{2\, i \, m \, z}{u})
\label{R.06}
\eeq
we re-write (\ref{R.04}) as
\beq
&& G(\phi \, t; \phi' \, t') = \sqrt{ \frac{I}{2 \, \pi \, i \, \tau}} \exp(-i \sigma_{3} \Delta \tau) \sum_{m} \times \nonumber \\
&& exp\Bigl (i I\frac{ (2 m \pi - \phi+\phi')^2}{2 \tau } \Bigr ) 
\exp(i \, \alpha (2 m \pi - \phi +\phi') \,\sigma_{3}  ). \nonumber \\
\label{R.07}
\eeq
In this form, the propagator contains products that are proportional to the time interval $\tau$, 
and are of a dynamical origin, with factors that are independent of $\tau$ and
have a geometric, or topological, origin. 
Consider the classical equation of motion for a free  rotor
$ \phi(t) = \omega \, (t-t') + \phi' $ or, if we set $t'=0$,  $\phi(t) = \phi'+ 2 \, m \pi $, for a rotor trajectory that
encompass $m$ circuits in a given
time period $\tau$. The resulting classical  action 
\beq
S_{m}(\tau) = \int_{0}^{\tau} \frac{I}{2} \omega^{2} = I \frac{(\phi -\phi' -2 m \pi)^{2}  }{2 \tau} 
\label{R.08}
\eeq
where we used the fact that $ \omega =  (\phi -\phi' -2 \, m\, \pi)/\tau $
Therefore,  
\beq
&& G(\phi \, t; \phi' \, t') = \sqrt{ \frac{I}{2 \, \pi \,  \, i \,\tau}} \sum_{m} \exp(i\, S_{m}(\tau)) 
 \exp(-i \sigma_{3} \Delta \tau) \times \nonumber \\
&&  \exp(i \sigma_{3} \alpha (2 m \, \pi - \phi+\phi')  ). 
\nonumber \\
\label{R.09}
\eeq
The partition function corresponds to the trace over all closed paths in which $ \phi=\phi' $, and the time interval $\tau$ is Wick rotated onto the imaginary
axis. With the replacement $ \tau \rightarrow - i \, \beta $, we obtain
\beq
&& {\cal Z}_{WY} = \int_{0}^{2 \pi} d \phi \, Tr  \, G(\phi,-i \, \beta,\phi,0) = \nonumber \\
&& 2 \sqrt{ \frac{I}{2 \, \pi \, \beta}} \cosh(\beta \Delta)  \sum_{m} \exp(-\frac{2 \pi^2 m^2 I}{\beta}) \cos(2 m \pi \alpha). 
\nonumber \\
\label{R.10}
\eeq
In the same manner we construct the partition function for $ x_{0} >1 $.
Thus, we find
\beq
&& {\cal Z}_{WY} = {\cal Z}_{0} \, \sum_{m} \exp(-\frac{2 \pi^2 m^2 I}{\beta}) \cos(2 m \pi \alpha)
 \quad  x_{0}  < 1   \nonumber \\
&&  {\cal Z}_{WY} =   {\cal Z}_{0} \, \sum_{m} \exp(-\frac{2 \pi^2 m^2 I}{\beta})   \quad  x_{0}  > 1  \nonumber \\
&& {\cal Z}_{0} \equiv 2 \sqrt{ \frac{I}{2 \, \pi \, \beta}} \cosh(\beta \Delta). 
\label{S11.c}
\eeq
In expression (\ref{S11.c}) the partition function is expressed as a product of a purely dynamical contribution
$ {\cal Z}_{0}$, and 
\beq
&& r_{\lessgtr}(\beta) \equiv 1 + \sum_{m=1}^{\infty} \exp(-\frac{2 \pi^2 m^2 I}{\beta}) \, 2 \cos(2 m \, \pi \, \alpha_{\lessgtr} ) \nonumber \\
&& \alpha_{\lessgtr} = \alpha \quad  x_{0} < 1, \quad \alpha_{\lessgtr} = 0 \quad x_{0} > 1
\label{S11.cc}
\eeq
which is modulated by a topological term $ \cos(2 m \, \pi \, \alpha_{\lessgtr} ) $ proportional
to the trace of the Wilson loop integral (\ref{S.09}) corresponding to winding number $m$.

In Fig (\ref{fig:fig5}) we plot the ratio $r_{<}(\beta)$ as a function of the inverse temperature $\beta$.
The graph illustrates significant variation of that ratio with respect to $\alpha$ at lower temperatures. 
\begin{figure}[ht]
\centering
\includegraphics[width=0.9 \linewidth]{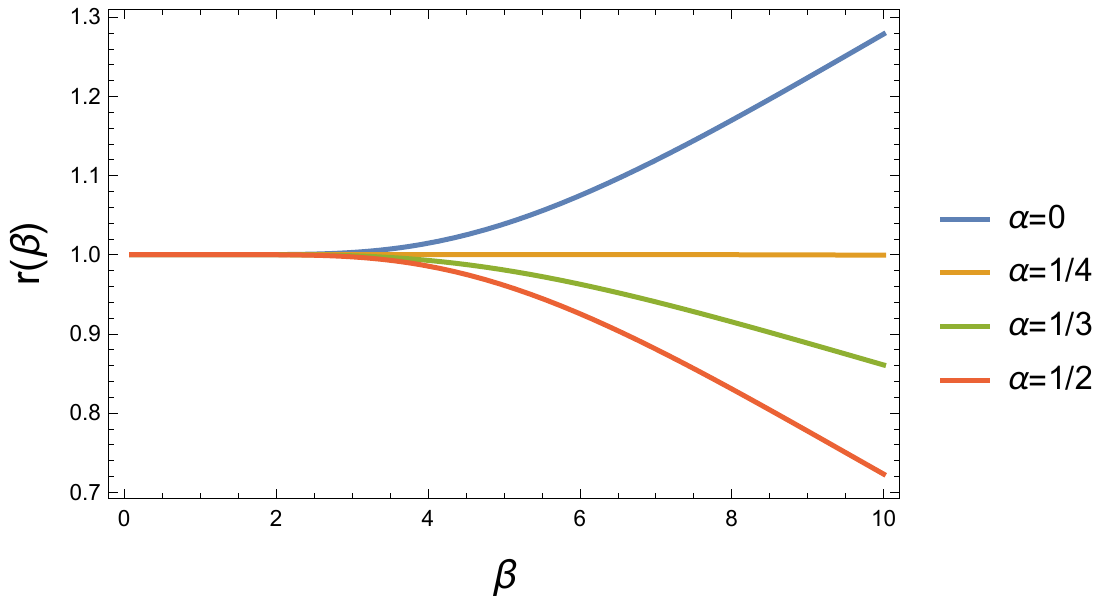}
\caption{\label{fig:fig5} Plot of the ratio $ r(\beta) \equiv {\cal Z}_{WY}/{\cal Z}_{0} $
as a function of $\beta$. The labeled curves correspond to different values of $\alpha$. We set the moment of inertia $I=1$. }
\end{figure}
For $x_{0}| >1 $, $r(\beta) $ undergoes a phase change as the curve is independent of variations in $ \alpha$, 
and reverts to that labeled by $\alpha=0$ in that figure.

It is now instructive to compare the behavior of the gauge invariant partition function for the Wu-Yang flux tube with that of 
the system described by the partition function
\beq
{\cal Z} = \sum_{m} \exp(-\beta E_{+}) + \sum_{m} \exp(-\beta E_{+}) 
\label{S11.d}
\eeq
where $E_{\pm} $ are given by expression (\ref{1.12}). The latter correspond to the partition function of our physical model; 
a neutral particle constrained on a rotor track in the presence of magnetic field (\ref{0.1}).

Instead of comparing $ {\cal Z}_{WY} $ with $ {\cal Z}$, we compare terms that only include the topological contribution to the partition functions. To that end we define
\beq
{\tilde r }(\beta) \equiv {\cal Z }/{\tilde {\cal Z}}_{0} 
\label{S11.e}
\eeq
where ${\tilde {\cal Z}}_{0}  $ is defined in (\ref{S11.c} ) but modified
by the contribution of the induced, scalar, counter term 
$$ \frac{\sin^{2}(\theta) }{8 I} = \frac{\alpha (1-\alpha)}{2 I} $$
introduced in Eqs. (\ref{3.08}) and  (\ref{3.11}), i.e. 
\beq
{\tilde  {\cal Z}}_{0} \equiv 2 \exp(-\beta \frac{(\alpha-\alpha^2)}{2 I}) \sqrt{ \frac{I}{2 \, \pi \, \beta}} \cosh(\beta \Delta).
\label{S11.f}
\eeq

\begin{figure}[ht]
\centering
\includegraphics[width=0.9 \linewidth]{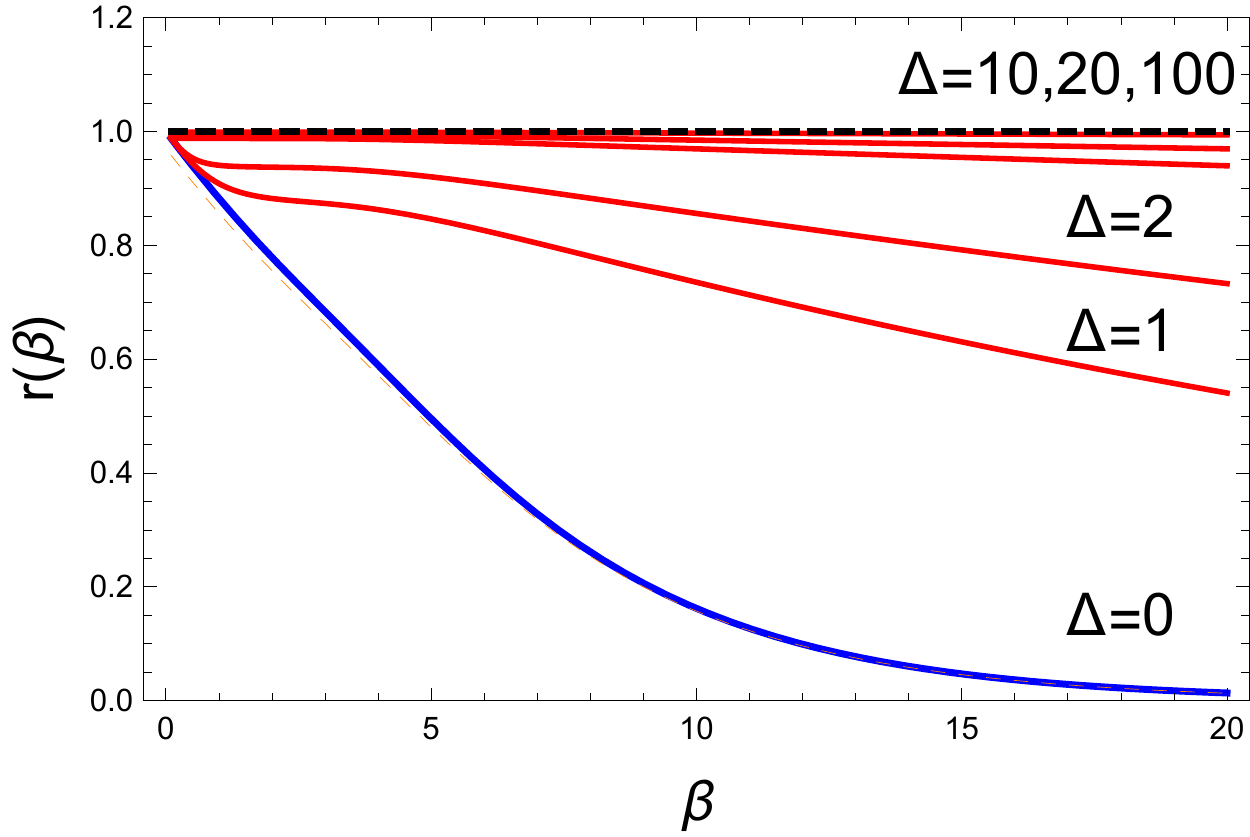}
\caption{\label{fig:fig6} Plot of the ratio $ {\tilde r}(\beta)$, for different values of the energy defect $\Delta$,
as a function of $\beta$. In this graph we chose the parameter values $\alpha=1/2, x_{0}=0, I=1$. The ratio
corresponding to the Wu-Yang system is given by the constant, black, dashed line. The red dashed line (superimposed by the blue line labeled $\Delta =0 $) corresponds to a free rotor.}
\end{figure}
In Fig. (\ref{fig:fig6}) we plot the ratio $ r(\beta) $ for the values $\alpha=1/2$, $x_{0}=0$,
as a function of the inverse temperature $\beta$ and the energy defect $\Delta$. The (blue) curve corresponding to energy defect
$ \Delta =0 $ is identical to the curve obtained for the partition function of a free rotor (i.e. without a non-trivial gauge couplings).
Since gauge potential (\ref{3.03}) describes a pure gauge it is plausible that it does not contribute the value of
the partition function ${\cal Z}$. However, for non-vanishing energy defects the graph shows a strong dependence of ${\cal Z}$
on the topological factor $ {\tilde {r } }$. For energy defect $ \Delta =100$, the value of  $ {\tilde {r } }$ is
almost identical to, at low temperatures ($\beta >> 1$), to the value predicted by the Wu-Yang flux tube given by
expression (\ref{S11.cc}) and shown by the dashed line in that figure. We conclude that for large values of $\Delta$
the gauge invariant partition function for the system defined in Eq. (\ref{1.01a}) approaches that of particle
coupled to Wu-Yang flux tube. Though gauge potential (\ref{3.03}) is that of a pure gauge, the energy defect $\Delta$
breaks a restricted spacial gauge symmetry as it corresponds to the time component of a $3+1$ gauge field\cite{zyg87a}. 
Consequently we find a non-trivial, non-Abelian, Wilson loop contribution to the partition function. If we restrict our attention to the ground state, the latter appears as an Abelian holonomy whose semiclassical analog (in which the quantum variable $\phi$ is demoted to a classical parameter $\phi(t)$ ) corresponds to Berry's geometric phase\cite{ber84,Cohen2019}.

Let's define amplitude $G$, so that
\beq
\psi = U \, \exp(-i \, \sigma_{3} \, \Delta \, t) G
\label{X0.1}
\eeq
where $U$ is defined in Eq. (\ref{3.01}). Inserting (\ref{X0.1}) into the time dependent version of Eq. (\ref{3.02}), we
obtain
\beq
-\frac{1}{2I} ( \partial_{\phi} - i {\bm A}(t))^{2}G = i \, \frac{\partial G}{\partial t}
\label{X0.2}
\eeq
where
\beq
&& {\bm A}(t) =
  \frac{1}{2} \left ( \begin{array}{cc}   \cos\theta -1 & i \, \sin\theta \exp(-i \phi(t)) 
  \\
-i \, \sin\theta \exp(i \phi(t))  & (1- \cos\theta) \end{array} \right )
\nonumber \\
&& \phi(t) \equiv \phi - 2 \, \Delta \, t.
\label{X0.3}
\eeq
$ {\bm A}(t) $, like ${\bm A}$ in Eq. (\ref{3.03}), is a pure gauge and
generates a trivial Wilson loop integral. However,
if we replace the off-diagonal
components of (\ref{X0.3}) with a time expectation value, over interval $\tau$,
$$  \pm i \, \sin\theta \, \langle \exp(\mp i \phi(t))  \rangle 
\approx  \pm i \, \sin\theta \langle \exp(\mp i \phi) \, {\cal O}(\frac{1}{\tau \,\Delta}) $$
which as $\Delta \rightarrow \infty$ we ignore. In this approximation pure gauge ${\bm A}(t)$ is
replaced with the gauge potential of a non-Abelian WY flux tube.

\subsection{Shifted magnetic vortex field}
In the previous section we demonstrated how, in the limit $\Delta \rightarrow \infty$,  the eigen-solutions to Hamiltonian 
(\ref{1.01a})) tend to those described by an effective
Hamiltonian containing a Wu-Yang flux tube.
Suppose we have the following ${\bm B} $ field configuration
\beq
{\bm B}/B_{0}  =   \, \frac{ - y  \,    {\bm {\hat i}}   }{ \sqrt{(x-x_{0})^2 + y^2}}  +  \frac{ (x-x_{0})   \,    {\bm {\hat j}}   }{ \sqrt{(x-x_{0})^2 + y^2} } 
\label{S.12}
\eeq
 which describes a vortex configuration centered at $ ( x=x_{0},y=0 ) $. The Hamiltonian in 
 (\ref{0.2}) 
 $ \mu \, {\bm \sigma}\cdot {\bm B} $ is given by
 \beq
&&  \left(
\begin{array}{cc}
 0 & \frac{i \Delta (x_{0}-x +i y)}{\sqrt{(x_{0} -x)^2+y^2}} \\
 -\frac{i \Delta  (x_{0} -x-i y)}{\sqrt{(x_{0} -x)^2+y^2}} & 0 \\
\end{array}
\right) \nonumber \\
&& \Delta = \mu B_{0} 
\label{S.13}
\eeq
and replacing $ x \rightarrow \cos\phi $, $ y \rightarrow \sin\phi $
the above expression can be re-written as
\beq
&& \left(
\begin{array}{cc}
 0 & -i \Delta \, e^{-i \Omega (\phi )} \\
 i \Delta \, e^{i \Omega (\phi )} & 0 \\
\end{array}
\right)    \nonumber \\
&& \tan \Omega = \frac{\sin\phi}{\cos\phi - x_{0}} 
\label{S.14}
\eeq
Now if we define the operator
\beq
U = \exp(-i \sigma_{3} \Omega/2 ) \,    \exp(i \sigma_{1} \pi/4  )          \,   \exp(i \sigma_{3} \Omega/2 ) 
\label{S.15}
\eeq
we find that
\beq
&& H = U \, H_{BO} \, U^{\dag} \nonumber \\
&& H_{BO} = \left(
\begin{array}{cc}
 \Delta  & 0 \\
 0 &  - \Delta. 
\end{array}
\right)  
\label{S.16}
\eeq
Forming the non-Abelian connection  $ A  \equiv i U^{\dag} \frac{\partial }{\partial \phi}  U$ we find,
\beq
&& A = 
\frac{1-x_{0} \cos\phi}{2 + 2 x_{0}^2 -4 x_{0} \cos\phi}
\left(
 \begin{array}{cc} -1 & i \, \exp(-i \Omega) \\ -i \, 
 \exp(i \Omega) & 1 \end{array} \right ) \nonumber \\
\label{S.17}
\eeq
The diagonal components of $A_{d}$ of $ A$ has the form,
\beq
A_{d} = \sigma_{3} \,  \frac{x_{0} \cos \phi -1}{2 x_{0}^2-4 \cos \phi  \, x_{0}+2} 
\label{S.18}
\eeq
and for the special case $x_{0}=0$ reduces to $ A_{d} = \sigma_{3}/2$ and  describes the non-Abelian
Wu-Yang flux tube of ``charge'' $1/2$ centered at the origin. In Fig. (\ref{fig:fig7}) we
plot, with the red solid lines, the energy spectrum calculated for Hamiltonian
(\ref{1.01a}) using field (\ref{S.12}) for values of $ x_{0}$ ranging from $x_{0}=0$ to $x_{0}=1.8$. Superimposed on the figure, by the blue dotted lines, is the corresponding spectrum for a rotor system minimally coupled to the gauge field of a Wu-Yang flux tube centered at $x_{0}$, and calculated using the analytic formulas given in Eqs. (\ref{S.10b}) and (\ref{S.11b}). The dashed blue lines correspond to the eigenvalues for a free planar rotor.

\begin{figure}[ht]
\centering
\includegraphics[width=0.9 \linewidth]{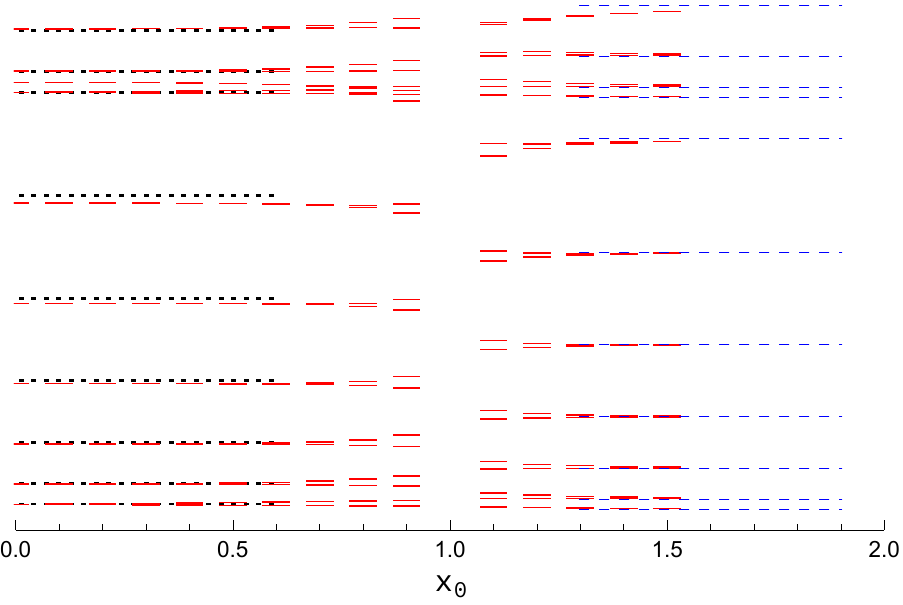}
\caption{\label{fig:fig7} Energy
spectrum for the rotor as a function of the vortex origin. For
$ x_{0} < 1 $, it is circumscribed by the rotor track.
Discontinuity at $ x_{0}=1$, demonstrates 
evidence of a topological phase transition. }
\end{figure}

\section{The Importance of being non-Abelian}

Consider the  Schr{\"o}dinger equation for a spin-1/2 particle of mass $m$
\beq
&& - \frac{1}{2 m} ( {\bm \nabla}  - i \, {\vec{\bm A'}})^2 \psi - {\bm A'}_{0}(\phi) \psi = i \frac{\partial \psi}{\partial t} \nonumber \\
&& { {\bm  A'}}_{0}(\phi) = \exp(-i {\bm a} \, \phi) {\bm \Delta} \exp(i {\bm a} \, \phi) \nonumber \\
&& {\bm \Delta} \equiv - \Delta \, { \sigma}_{3}, 
\label{0.00}
\eeq 
where ${\bm a}$ is a constant $2 \times 2 $ hermitian matrix, $ \phi$ is the azimuthal angle in a cylinderical coordinate system and $ {\vec {\bm A'}}, {\bm A'}_{0} $ are the spatial and time components of a 3+1 matrix-valued (i.e. non-Abelian) gauge potential. Let ${\vec {\bm A'}}=0 $, and so Eq. (\ref{0.00}) describes a spin-1/2 particle 
coupled to a matrix, or spin-dependent, scalar potential $-{\bm A'}_{0}$. With gauge transformation
$ \psi = {\bm U} \,  F $, $ {\bm U} =\exp(-i {\bm a} \, \phi )$, amplitude $F$ obeys,
\beq
- \frac{1}{2 m} ( {\bm \nabla}  - i \, {\vec{\bm A}})^2 F - {\bm A}_{0}(\phi) \, F = i \frac{\partial F}{\partial t} \nonumber \\
\label{0.01}
\eeq
where
\beq
&& {\vec {\bm A}} = U^{\dag} {\vec {\bm A'}} U + i \, U^{\dag} {\bm \nabla} {\bm U}  = \frac{{\bm a}}{\sqrt{x^2+y^2} }  \, {\bm {\hat \phi}}
\nonumber \\
&& {\bm A}_{0} = U^{\dag} {\bm A'}_{0} U + i \, U^{\dag} \frac{\partial U}{\partial t} = {\bm \Delta}
\label{0.01a}
\eeq
The similarity of Eq. (\ref{0.00}) with (\ref{0.01}) is a reflection of the fact that the Schr{\"o}dinger equation is covariant, or form invariant, with respect to gauge transformations. Observable quantities, the eigenvalues of operators, are gauge
invariant.

Now gauge transformation ${\bm U}(\phi) = \exp(-i {\bm a} \phi )$ must be single-valued, i.e. ${\bm U}(0) = {\bm U}(2 \pi)$, and so ${\bm a}$
has the form 
\beq
 {\bm a} = {\bm Z}^{\dag} \left ( \begin{array}{cc} m & 0 \\ 0 & n \end{array} \right )  {\bm Z} 
\label{0.01b}
\eeq
where $m,n$ are integers, $\theta, \gamma$ are constants, and 
$${\bm Z} = \exp(i \theta \sigma_{2}/2) \exp(- i \gamma \sigma_{3} /2) $$
is a constant unitary matrix. For the sake of simplicity, we consider the case 
$n=-m $ and so 
\beq
{\bm a} = q \, \left ( \begin{array}{cc} \cos\theta  & \exp(i \gamma) \sin\theta  \\
 \exp(-i \gamma) \sin\theta  & -\cos\theta \end{array} \right ), 
\label{0.02}
\eeq
where $q$ is an integer and $\theta, \gamma$ are parameters, satisfies Eq. (\ref{0.01b}). A full quantum description of this model is given 
in Appendix A, but here we first explore the behavior of the Wilson loop integral of the 3+1 gauge potentials ${\bm a}, {\bm A}_{0}.$

Consider the following path-ordered Wilson-loop integral,
\beq
&& {\bm W}(C_{0}) = P \, \exp(i \int_{C_{0}} d{\vec {\bm R}} \cdot {\vec {\bm A}} )  = \nonumber \\
&& \exp(i {\bm a} \int d\phi) 
\label{0.02a}
\eeq
where we used ${\vec {\bm A}} $ defined in Eq. (\ref{0.01a}), $ C_{0}$ is a closed path that circumscribes the origin in the, $z=0$,  $ xy $ plane
and $d \phi $ is the differential angle, with respect to the origin, of a segment of an arc along the path. Since $\int d\phi = 2 \pi m$, where $m$ is the winding number of the path, 
\beq
&& {\bm W}(C_{0}) = \exp(i \, {\bm a} \, 2\pi \, m)  = \nonumber \\
&& {\bm Z}^{\dag} \exp(i \, 2 \pi \, q \, m  \, \sigma_{3} )\, {\bm Z} = \mathbb{1}.
\label{0.02b}
\eeq 
This identity is simply a reflection of the fact that $ {\vec {\bm A }} $ is a pure gauge. 
\subsection{Wilson line in space-time.}
In our discussion so far we noted that the partition function of our spin-1/2 systems
contain Wilson loop contributions that arise from non-trivial
gauge fields, despite the fact that the spatial components ${\vec{\bm A}} $ of the 3+1 gauge potentials describe a pure gauge. 

To achieve a better understanding of how non-trivial Wilson loop contributions arise in systems that are putatively coupled to a pure gauge, we note that in evaluation of the partition function we need to take into account paths in space and time.  Therefore, we consider a general path integral along an arbitrary path (not including the origin) $C(a,b)$ from point
$a$ to $b$ for  gauge field $ {\bm A}_{\mu}$. Here $\mu$ is an index that identifies  a space-time component
\beq
 {\bm A}_{u} = \{ {\bm A}_{0},{\bm A}_{x},{\bm A}_{y}, {\bm A}_{z} \}  \quad \mu =0,1,2,3 
\nonumber 
\eeq
and we use a summation convention so that
\beq  
&& {\bm W}(a,b) \equiv P \exp(i \int_{C(a,b)}  \,  dz^{\mu} {\bm A}_{\mu} ) \nonumber \\
&& dz^{\mu} {\bm A}_{\mu} \equiv  dt \, A_{0} + {\vec {\bm A}}\cdot d {\vec R}
\label{0.02bb}
\eeq

With gauge transformation $ \psi = {\bm U} \psi' $, the gauge potentials\cite{Mak09} 
\beq
&& {\bm A}_{\nu} \rightarrow { {\bm A}'}_{\nu}  = {\bm U}^{\dag} \,{\bm A}_{\nu} {\bm U} + i \, {\bm U}^{\dag} \partial_{\nu} {\bm U}, \quad \nu =1,2,3\nonumber \\
&& {\bm A}_{0} \rightarrow {\bm A}'_{0} = {\bm U}^{\dag} \,{\bm A}_{0} {\bm U} + i \,{\bm U}^{\dag} 
\partial_{t} {\bm U} \nonumber \\
&& {\bm W}(a,b) \rightarrow {\bm U}^{\dag}(b) \,{\bm W}(a,b)  \, {\bm U}(a).
\label{0.02bc}
\eeq
Consider paths of the type illustrated in Fig. (\ref{fig:fig8}). They are trajectories in a manifold that is a Cartesian product of the coordinates in the  $xy$ plane with a 1-dimensional manifold labeled by time $t$. 
 The trace of ${\bm W}(a,b)$ for an open-ended path is not, 
 in general, gauge invariant. However, we evaluate the integral only along paths in which the projection of coordinates $a,b$ onto the spatial plane are equal at the initial and final points of the trajectory. We also limit the gauge group to time independent gauge transformations ${\bm U}$
 so that the trace of $ {\bm W}(a,b)$ is invariant under this group of transformations. Below we study the properties of $ {\bm W}(a,b)$  as a function of the defect parameter $\Delta$.
 
\begin{figure}[ht]
\centering
\includegraphics[width=0.9 \linewidth]{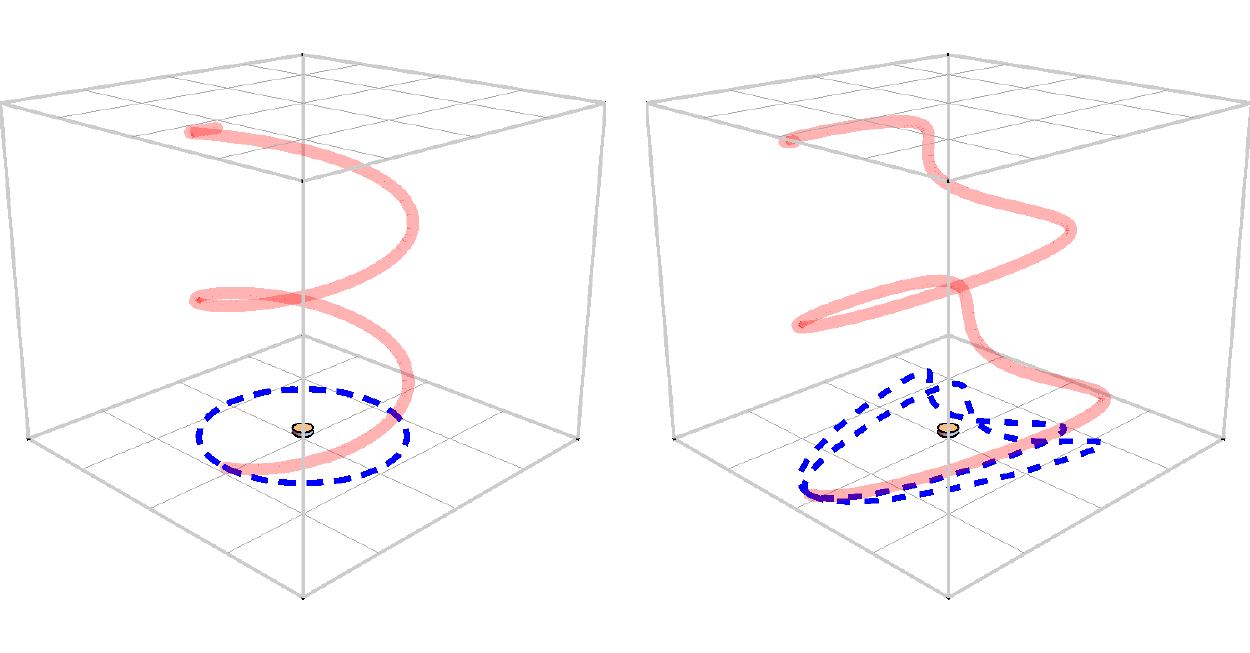}
\caption{\label{fig:fig8}  Space-time paths for the Wilson line integral
(\ref{0.02bb}). Deformations within a set of projected paths on the $xy$ plane,
shown by the dashed lines,
that share the same winding number about the origin do not alter the value 
of the integral.}
\end{figure}

We parameterize the trajectory $z(\tau)$ 
\beq
{\bm z}(\tau) = x(\tau) {\hat {\bm i}} +  y(\tau) {\hat {\bm j}} + f_{t}(\tau) \, {\hat {\bm k}}
\label{0.03}
\eeq
where $  {\hat {\bm i}}, {\hat {\bm j}} $ are the basis vectors in the spatial plane, and $ {\hat {\bm k}}$
is the unit vector orthogonal to that plane and which we take to define the time axis, so that the physical time $t \equiv f_{t}(\tau)$. The functions $x(\tau),y(\tau), f_{t}(\tau)$ are arbitrary but satisfy the conditions
$ x(0) = x(t_{f}), y(0) = y(t_{f})$ where $ 0 < \tau \leq t_{f}$ in order
for the path to make a closed loop, in the $xy$ plane at $\tau = t_{f} $. Using Eqs. (\ref{0.01a},\ref{0.02bb},\ref{0.03})
we get
\beq
&& {\bm W}(a,b) =  P \, \exp\Bigl (i \int_{C(a,b)}  d{z}^{\mu}  {\bm A}_{\mu} ) = \nonumber \\
&& T \exp(i \int_{0}^{t_{f}}
\, d\tau \, ( \frac{d \phi}{d \tau} \, {\bm a}    + 
\frac{d f_{t}}{d \tau} {\bm \Delta} ) \,  \Bigr ) 
\label{0.05}
\eeq
where $T $ denotes time-ordering. If ${\bm a}$ commutes with ${\bf \Delta} $ expression (\ref{0.05})
factors into a product of the trivial Wilson-loop integral (\ref{0.02a}) and a dynamical contribution generated by ${\bm A}_{0}$.
For gauge potential (\ref{0.02}) such a factorization is not possible, as $[{\bm a}, \sigma_{3}] \neq 0$. However, integral (\ref{0.05} ) can easily be evaluated for a class of paths where $ d \phi/{dt} \equiv \omega  $ is constant. Since
$$ \frac{d \phi}{d t} = 
\frac{d \phi}{d \tau} \frac{1}{f'_{t}(\tau)} $$
and $\phi(t_{f})=\phi(0) $ ${\rm Mod}(2 \pi)$
\beq
&& {\bm W}(a,b) =
 T \exp\Bigl (i \int_{0}^{t_{f}}
\, d\tau \, \frac{d \phi}{d \tau} \, ( \, {\bm a}  
+ \frac{\bm \Delta}{\omega} ) \Bigr ) = \nonumber \\
&& \exp\Bigl (i \, 2 \pi\,  m ({\bm a}+ \frac{\bm \Delta}{\omega} ) \Bigr ), 
\label{0.06a} 
\eeq
where $m$ is the winding number of the path.
Exponentiation of expression (\ref{0.06a}) 
results in
\beq
&&  {\bm W}(a,b) = \nonumber \\
&& \cos(2 \pi m \, \Omega) \, \mathbb{1} -
i \frac{(\Delta/\omega - q \, \cos\theta)}{ \Omega}
\sin(2 \pi m \, \Omega) \, \sigma_{3} + \nonumber \\
&& i \frac{\, q \, \sin\theta}{\Omega} \, 
\sin( 2 \pi m \, \Omega)
(\cos\gamma \, \sigma_{1} - \sin\gamma \, \sigma_{2} ).  \label{0.06b}
\eeq
where 
\beq 
\Omega = \frac{\sqrt{\Delta^2 + q^2 \omega^2 - 2 \Delta \, q \, \omega \cos\theta}}{\omega}. 
\label{0.06c}
\eeq 

Let's define an effective vector potential
\beq
{\cal A}_{eff} \equiv \frac{\hat{\bm \phi} }{\sqrt{x^2 +y^2} } 
\Bigl ( {\bm a} + \frac{\bm \Delta}{\omega} ). 
\label{0.08}
\eeq
Unlike the pure gauge $ {\vec {\bm A} }$, defined in Eq. (\ref{0.01a}), $ {\cal A}_{eff} $ engenders a non-trivial Wilson loop integral ${\cal W}_{C} $ for any loop,
in the $z=0$ plane, enclosing the origin. Indeed, 
\beq
{\bm W}(a,b) = {\cal W}_{C} = \oint_{C} d {\vec {\bf R}} \cdot {\cal A}_{eff}
\label{0.09}
\eeq
where $C$ is the projection of the space-time path (\ref{0.03}) onto the $xy$ plane. Because $x(0)=x(t_{f}), y(0)=y(t_{f})$,
$C$ forms a closed loop.

In summary, we  demonstrated how the space-time open-ended path integral of a $3+1$
non-Abelian gauge potential leads to a non-trivial
Wilson loop integral of an effective gauge field 
${\cal A}_{eff}$. For time independent
gauge transformations, the trace of ${\bm W}$ is gauge invariant. As ${\cal W}_{C}$ depends only on the winding number, $ C$ can be shrunk to an 
infinitesimal loop about the origin without altering the
value of ${\cal W}_{C}$. Thus ${\cal A}$
represent the gauge potential of a Wu-Yang flux tube of "charge"
$ \pm \Omega $, 
the eigenvalues of $ {\bm a} + \frac{\bm \Delta}{\omega}. $
In general,
${\bm W}(a,b)$ is a function of the dynamical parameters, $\Delta, {\bm \omega}$, but for large $ \frac{\Delta}{\omega} >>1 $, it tends to the product
\beq
&& {\bm W}(a,b) \rightarrow  \exp(-\frac{2 \, i \, m \pi}{\omega} \, \sigma_{3})
\exp(2 \, i \, m \, \pi \, q \, \cos\theta \, \sigma_{3} )
+ \nonumber \\
&& {\cal{O}}(\omega/\Delta )
\label{0.10}
\eeq

We evaluated ${\bm W}$ in
the adiabatic gauge\cite{zyg87a}, wherein ${\bm A}_{0}$ is diagonal. Because
 $ {\bm U}(\phi=0) = {\bm U}(\phi=2 \pi) = \mathbb{1}$,
${\bm W}$ is invariant under a gauge transformation
into the diabatic gauge\cite{zyg87a}.
The latter corresponds to Schr{\"o}dinger Eq. (\ref{0.00})
in which the spatial component ${\bm A'}=0$.
In that gauge
\beq
&& {\bm W}(a,b) =  P \, \exp(i \int_{C(a,b)} d{\bm z}^{\mu}  {\bm A'}_{\mu}  ) = \nonumber \\ 
&& T \exp \Bigl ( i \int_{0}^{t_{f}} d\tau \, 
\frac{df_{t}}{d \tau }
{\bm A'}_{0}(\phi(\tau)) \Bigr )= 
\nonumber \\
&& T \exp\Bigl (i \int_{0}^{\frac{2 \pi m}{\omega} } d t \, \exp(i {\bm a} \, \omega \, t) {\bm \Delta} \exp(-i {\bm a} \, \omega \, t) \Bigr )
\label{0.11}
\eeq
where we used the fact that $f_{t}(\tau) = t$ and 
$ \phi(t) = \omega \, t. $
Replacing the upper limit in integral (\ref{0.11}) with an
arbitrary time value $t$, we find that ${\bm W}(t)$ obeys 
a time dependent Schr{\"o}dinger equation.
\beq
&& i \, {\dot {\bm  W} }(t)  =
{\cal H}(t) \, {\bm W}(t) \nonumber \\
&& {\cal H}(t) = \exp(i {\bm a} \, \omega \, t) {\bm \Delta} \exp(-i {\bm a} \, \omega \, t).
\label{0.12}
\eeq
It can be integrated to give
\beq
{\bm W}(t) = \exp(-i\, {\bm a} \, \omega \, t) \exp \Bigl (i \,\omega \,t \,({\bm a} +
\frac{\bm \Delta}{\omega} ) \Bigr ).
\label{0.21a}
\eeq
Thus,
$$ {\bm W}(a,b) = {\bm W}(t = \frac{2 \pi m}{\omega} ) $$
where we used the fact that 
$ \exp(2\, i\, \pi \, m \, {\bm a}) = \mathbb{1}$.
In the adiabatic limit as $\omega \rightarrow 0$,
${\bm W}(a,b)$ tends to the limit Eq. (\ref{0.10}). 
In that expression, the first, dynamical, factor 
$ \exp(-\frac{2 \, i \, m \pi}{\omega} \, \sigma_{3}) $ depends on the length of time 
$ t_{f}=2 \pi m/\omega $ that it takes for the system to travel
from starting to end points. The second factor
$$ \exp(2 \, i \, m \, \pi \, q \, \cos\theta \, \sigma_{3} )$$ 
depends on spatial path taken. This factorization is in harmony with
the adiabatic theorem\cite{ber84}.

\section{On the Wilczek-Zee Mechanism}
In the previous sections we illustrated how non-trivial gauge structures arise in a vector space that is a direct product
of a two-state (or qubit) system with the Hilbert space of a  rotor.
It is straightforward to extend this formalism to systems possessing additional internal
degrees of freedom (e.g. spin-1 etc.). Indeed, this procedure is ubiquitous in theoretical studies of
slow atomic collisions and non-adiabatic molecular dynamics. In those applications it is especially applicable
if the total system energy $E << \Delta $ where $\Delta$ is an energy
defect that separates a sub-manifold of Born-Oppenheimer (BO) states separated by a large energy gap 
from energetically higher lying BO states.
Thus the Hilbert space amplitude is projected to a set of effective, or matrix-valued,
amplitudes in the sub-space. The resulting set of coupled equations constitute the Born-Huang\cite{BH}
approximation, or the method of Perturbed Stationary States\cite{mott49} (PSS). The latter
typically result in effective, non-trivial, non-Abelian gauge couplings among the sub-space amplitudes.

In a quasi-classical version of this procedure, Wilczek and Zee demonstrated how the projected
amplitudes, for a sub-manifold of degenerate energy eigen-states, acquire a non-Abelian
geometric phase during adiabatic evolution.

Below we consider a spin-1 rotor system in which two internal states
posses degenerate energy eigenvalues that are separated from the remaining
internal states by a large energy gap $\Delta$. To illustrate this mechanism we choose a straightforward extension of
Hamiltonian (\ref{0.00}) 
\beq
&& H = -\frac{1}{2 \, I} \partial^{2}_{\phi} + {\bm U}(\phi) \bm {V} {\bm U}^{\dagger}(\phi) \nonumber \\
&& {\bm U} = \exp(- i \, {\bm a}\, \phi)  \nonumber \\
&& {\bm a} = \left(
\begin{array}{ccc}
0 & \frac{\sin \theta }{\sqrt{2}} & -\frac{\sin \theta }{\sqrt{2}} \\
\frac{\sin \theta }{\sqrt{2}} & \cos \theta  & 0 \\
-\frac{\sin \theta }{\sqrt{2}} & 0 & -\cos \theta  \\ 
\end{array} \right ) \nonumber \\
&& {\bm V} = \left ( \begin{array}{ccc}
\Delta & 0 & 0\\
0 & e_{g}  & 0 \\
0 & 0 & e_{g}  \\ 
\end{array} \right )
\label{Z0}
\eeq
This particular choice for ${\bm a} $ guarantees that ${\bm U}$ is single-valued.  
For our purposes it is convenient to choose $e_{g}=-\sin^2\theta/4I $.

Defining the adiabatic gauge
amplitude $F$, so that 
$$ \psi = {\bm U} \, F $$ 
we obtain the matrix-valued Schr{\"o}dinger equation
\beq
-\, \frac{1}{2 I} (\partial_{\phi} - i\, {\bm a})^{2} \, F + {\bm V} \, F = i \frac{\partial F}{\partial t} 
\label{Z1}
\eeq
With ansatz
\beq
F = \exp(i m \phi)\exp(-i \, E \, t)  \, {\bm c} 
\label{Z2}
\eeq
where $ {\bm c}$ is a constant column matrix, we are led
to the eigenvalue equation  $ {\rm det } \, |{\bm h} - \mathbb{1} \,  E | =0 $ where
\beq
{\bm h} = \frac{ (\mathbb{1} \, m  -{\bm a})^2}{2 \,I}  + {\bm V} 
\label{Z3}
\eeq
Finding the eigenvalues of ${\bm h}$ involve solving for the roots of a cubic 
equation and for which analytic expressions, the Cardano formula, is available.
The latter can be used to construct the gauge invariant partition function
\beq
{\cal Z} \equiv \sum_{i=1}^{3} \sum_{m} \, \exp(-\beta \, E^{i}_{m}).
\label{Z4}
\eeq
to the required degree of accuracy.
The sums extend over the spectrum of ${\bm h}$, which are itemized by
the motional quantum number $m$, as well as the internal state quantum number $i$.  Here $\beta$ is an inverse temperature.

Instead, because $ \Delta >> e_{g}$, we use  the PSS approximation in which the amplitude
$ F$ is projected to a Hilbert subspace. In this case, the subspace is spanned by the degenerate eigen-states of ${\bm V}$, or
the computational basis for a single qubit.
Introducing the projection operator 
$$ P \equiv \left ( \begin{array}{ccc}
0 & 0 & 0\\
0 & 1  & 0 \\
0 & 0 & 1  \\ 
\end{array} \right ) $$
defining $$ G = P \, F $$ we obtain the PSS equations
\beq
&& -\, \frac{1}{2 I} (\partial_{\phi} - i \, {\bm a}_{p} )^{2} \, G + \nonumber \\ 
&& P \, \frac{ {\bm a}\cdot {\bm a} - {\bm a}_p \cdot {\bm a}_{p} }{2 I} \, G
+ {\bm V}_{p}  \, G = i \frac{\partial G}{\partial t}  \label{Z5}
\eeq
where $ {\bm a}_{p} \equiv P {\bm a} P, \, {\bm V}_{p} = P \,{\bm V} \,P $.  In this approximation we ignore
couplings between the $P$ and $ Q=\mathbb{1}-P $ sub-manifolds.

Though $V_{p}$ is diagonal and degenerate, the higher
order induced scalar term\cite{zyg86,zyg90} 
\beq P\, \frac{ {\bm a}\cdot {\bm a} - {\bm a}_p \cdot {\bm a}_{p} }{2 I} \, P = \nonumber \\
\frac{1}{4 I} \left ( \begin{array}{ccc}
0 & 0 & 0\\
0 & \sin^2\theta & -\sin^2\theta \\
0 & -\sin^2\theta  &  \sin^2\theta \\ 
\end{array} \right ) 
\label{Z6}
\eeq
is not.  An additional gauge transformation in the projected qubit subspace $ G = {\bm W} { G'} $ results in 
\beq
&&  -\, \frac{1}{2 I} (\partial_{\phi} - i \, {\bm a'}_{p} )^{2} \, G' + {\bm V'}_{p}  \, G' = i \frac{\partial G'}{\partial t}  \nonumber \\
&&  {\bm a'}_{p} = \cos\theta \, \sigma_{1} \nonumber \\
&& {\bm V'}_{p} = \frac{\sin^2\theta}{4 I} \, \sigma_{3} \nonumber \\
&& {\bm W} = \exp(- i \sigma_{2} \pi/4)
\label{Z7}
\eeq
where $ \sigma_{i}$ are the standard spin-1/2 Pauli matrices.

Because the eigen-states of ${\bm V'}_{p}$ are not degenerate, Eq. (\ref{Z7}) is no longer covariant under a Wilczek-Zee gauge transformation. 
In the latter formulation $\phi(t)$ is treated as a classical variable undergoing adiabatic evolution. Here 
$\phi$ is a quantum variable, and the symmetry responsible for the degeneracy in a quasi-classical formulation is broken.  However we can, as described in the previous sections, enlarge the gauge group by allowing the (matrix) scalar potential
to be treated as the time component of a $3+1$ gauge potential. 

Consider the gauge potential. 
\beq
{\bm A'}_{p} = \sigma_{1} \cos\theta \, \Bigl ( \frac{-{\bm {\hat i}}  \, y }{x^2+y^2}  +   \frac{{\bm {\hat j}} \, x }{x^2+y^2} \Bigr )
\label{Z8}
\eeq
which begets ${\bf a'}_{p}$ in Eq. (\ref{Z7}).
Its Wilson loop integral for a path $C$ circumscribing the origin
assumes the value
\beq
&& {\cal W}_{C}(m) \equiv Tr \, P \exp(i \oint_{C} \, d{\bm r} \cdot {\bm A
'}_{p} ) = \nonumber \\
&& 2 \cos( 2 \pi \, m \, \cos\theta ) 
\label{Z9}
\eeq
where $m$ is the winding number. For values $\cos\theta  \notin \mathbb{Z} $, identity (\ref{Z9}) demonstrates that
$ {\bm a'}_{p}$, unlike ${\bm a}$ in Eq. (\ref{0.01b}),  is not a pure gauge. The energy eigenvalues associated with Eq. (\ref{Z7}) are 
\beq
&& E = e_{0} \pm e_{1}  \nonumber \\
&& e_{0}= \frac{m^{2} + \cos^2\theta }{2 \, I} \quad e_{1} = \frac{1}{I} \sqrt{ \frac{\sin^4\theta}{16} + m^2 \, \cos^2 \theta}
\label{Z10}
\eeq
and so the reduced partition function
\beq
{z} =2  \sum_{m}  \exp(-\beta \, e_{0}) \cosh(\beta \,e_{1}). 
\label{Z11}
\eeq
For higher temperatures, or $ \beta << 1$, we can approximate
$$ e_{1} \approx \frac{| m |}{I} \, \cos\theta $$
which implies that
\beq
z \approx 2 \sum_{m} \,  \exp(- \beta \, \frac{m^{2} + \cos^2\theta }{2 \, I} ) \cosh(\beta \, \frac{m \, \cos\theta}{I})
\label{Z12}
\eeq
Applying a Poisson transformation, we get
\beq
&& z \approx 2 \sqrt{\frac{2\pi \, I}{\beta} } 
\sum_{m} \exp(-\frac{2 \, I \, m^2 \pi^2}{\beta}) \cos(2 \pi \, m \, \cos\theta) \nonumber \\
\label{Z12a}
\eeq
In order to obtain the total partition function $ \cal{Z}$, we must include the contribution from the distant state whose energy eigenvalue
$ E^{i=1}_{m}  >> e_{0} \pm e_{1}$. In solving for the eigenvalues of ${\bm h}$ we find that
\beq
E^{i=1}_{m} = \frac{m^2}{2 \, I} + \frac{\sin^2 \theta/2}{2 \, I} + \Delta + {\cal O}(\frac{1}{\Delta}) + \dots
\label{Z13}
\eeq
and so the leading order contribution is dominated by the term $\exp(-\beta \, \Delta ) \rightarrow 0$ as $ \Delta \rightarrow \infty$.
Therefore,
\beq
{\cal Z} \approx  {z} = \sqrt{\frac{2\pi \, I}{\beta} }  \sum_{k}  \exp(-S_{0}(k)) \cos({\cal W}_{C}(k))
\label{Z15}
\eeq
where $ S_{0}(k) ={2 \pi^2 k^2 I}/\beta  $ is the classical action for a free rotor making $k$ complete circuits in a given time interval. It contains
a dynamical contribution, proportional to the classical action, that is modulated by a purely topological term, the Wilson loop
integral ${\cal W}_{C}(k)$. At higher temperatures $z$ is largely dominated by
contributions from the classical action 
and so we investigate the behavior of $ \cal{Z} $ in the low temperature $ \beta \rightarrow \infty $ limit. A detailed derivation is given in Appendix B and according to Eqs. (\ref{A0.5}) \beq
&& z \rightarrow  2\sqrt{\frac{2 \, \pi I}{\beta}} \,\exp(\beta \, V(\theta) ) \times \nonumber \\
&& \sum_{k}   \exp(-S_{0}(k)) 
\cos(2 \, \pi \, k \, \Omega )
\label{Z.16}
\eeq
in that limit. $ S_{0}(k)$ is the Wick rotated action for a free rotor undergoing $k $ circuits
and
\beq
&& V(\theta) = -\frac{\cos^2\theta}{I} + \alpha_{0} + \frac{ {I \, \alpha_{1}}^2}{2} \nonumber \\
&& \Omega =  -\frac{\sin^2\theta}{4} + \sqrt{\frac{\sin^2\theta}{16} + \cos^2\theta }.
\label{Z.17}
\eeq
In Fig. (\ref{fig:fig9}) we plot
$$  
-\frac{\partial \ln z}{\partial \beta } 
$$
as $ \beta \rightarrow \infty $, and which represent the ground state energy. The solid
line denotes the ground state energy for Hamiltonian
(\ref{Z0}), the dashed line the adiabatic energy 
$ e_{g}$, and the circle icons denote energies 
obtained in the PSS approximation and calculated using
expression (\ref{Z.16}) for the partition function. The latter
approximation is accurate for values $ \Delta/e_{g} >>1. $ According to expression (\ref{Z.16}), the term $m \, \Omega$ is independent of the temperature
parameter $\beta$ and is therefore of topological origin. The cross icons in that figure
represent the energies obtained by artificially setting $\Omega=0$ in expression (\ref{Z.16}). The difference between those values
and the ones laying on the solid line, underscores the significance of that topological contribution. Interestingly, unlike in the high temperature limit, the value for $2 \pi \, m \, \Omega$ does not equal the Wilson loop integral ${\cal W}_{C}$ of the projected gauge potential $ {{\bm A}'}_{p}.$

\begin{figure}
    \centering
    \includegraphics[width=0.8 \linewidth]{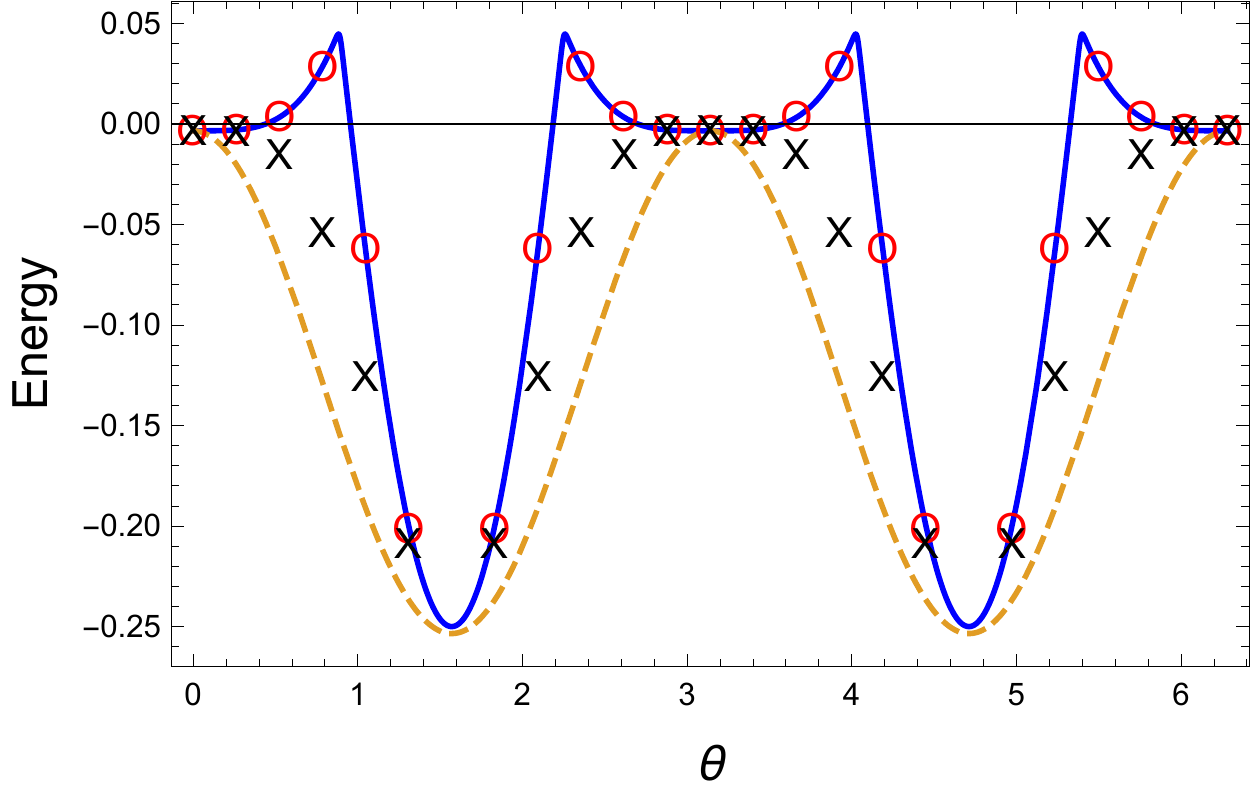}
    \caption{Ground state energy of Hamiltonian (\ref{Z0}) as a function of gauge parameter $\theta$. }
    \label{fig:fig9}
\end{figure}

\section{Summary and Discussion}

The gauge principle forms a cornerstone to our modern understanding of the fundamental constituents of matter. Quantum Electrodynamics (QED) is the best known example
of an Abelian gauge theory, and its non-Abelian generalization illuminates the landscape within the nucleus.

Gauge invariance guarantees charge conservation, and is the guiding principle that insures a gauge field's raison'detre. For example, the
following Hamiltonian (up to a surface term) for a scalar field $\phi$ 
\beq
{\cal H} = \frac{-1}{2m} \int d^{3} {\bm x} \, 
\phi^{\dag}(\bm x) {\bm \nabla }^2  \phi({\bm x})
\label{U0.00}
\eeq
is not invariant under the replacement of 
field operator $ \phi({\bm x}) $  
with $ \exp(i \Lambda({\bm x})) \phi({\bm x}).$ Introducing 
an auxiliary quantum field ${\bm A}$ so that
\beq
&& {\cal H} =
\frac{-1}{2 m} \int d^{3} {\bm x} \, \phi^{\dag}(\bm x) 
({\bm \nabla} - i {\bm A})^2 \phi({\bm x})
\label{U0.01}
\eeq
gauge invariance is enforced provided that as 
$ \phi({\bm x})  
\rightarrow \exp(i \Lambda({\bm x})) \phi({\bm x})$,  
$ {\bm A} \rightarrow {\bm A} + {\bm \nabla} \Lambda $.

In quantum mechanics (QM) the Schr{\"o}dinger equation is not invariant under a gauge 
transformation of the wave amplitude, however the eigenvalues of operators, i.e. observables, are. Dirac\cite{Dirac31} argued that a Schr{\"o}dinger description in which the wave function is minimally coupled to a gauge potential is equivalent to a gauge field free
theory whose wave amplitudes posses non-integrable\cite{Dirac31,wuyang75}, or Peirls\cite{spiel12} phase factors. 

In this paper we provided examples
of pedestrian quantum systems in which gauge structures arise in a natural manner without the need to summon the former. This feature of QM has long been noted in studies of atomic and molecular systems\cite{mead76,moo86,zyg86,zyg87a}. But, as those descriptions require the application of Born-Oppenheimer like approximations, predictions are open to interpretations that attracts skepticism\cite{Gross}. For example, laboratory searches for the Molecular Aharonov-Bohm Effect (MAB)\cite{mead80}, in the reactive scattering of molecules, has had a long and controversial history\cite{zyg2016,kendrick18,Yuan1289}. In this paper we addressed some of those concerns in two ways, (i) we identified systems that allow analytic solutions, and (ii) explicitly demonstrated the dependence of gauge invariant quantities (e.g. the partition function) on the Wilson loop integral of a non-trivial  gauge potential. Furthermore, our analysis did not require the semi-classical notion of adiabaticity, or degeneracy in the adiabatic eigenvalues.
Unlike gauge quantum fields, quantum mechanical gauge potentials, discussed here, do not exhibit dynamic content (but see Appendix C).

In the remaining discussion we address possible laboratory demonstrations of effects predicted and discussed in this paper. Though we are unable to comment on the viability of present day laboratory capabilities to realize the double slit system discussed in the introduction, we anchor our
focus on recent laboratory efforts 
to simulate a coherent quantum rotor. For example,
a planar quantum rotor was  simulated\cite{Urban}
in a cylindrical symmetric ion trap in which a pair of $^{40} {\rm Ca}^{+} $ ions formed a two-ion Coulomb crystal. That experiment demonstrated a capability to prepare and control angular momentum states.   
Along those lines we propose trapping a spin - 1/2 ion in a 
toroidal trap as shown in Fig (\ref{fig:fig10}). In that figure a positively charged spin-1/2 ion, such as $ {\rm Ca}^{+}$ in its ground state, is trapped in the torus. Instead, one can also consider a pair 
ions forming a Coulomb crystal, as described in \cite{Urban}. The latter simulates, after factoring out the center of mass motions, a single ion rotor. However, for the sake of illustration, we limit this discussion to a 
toroidal trap configuration. 
\begin{figure}[ht]
\centering
\includegraphics[scale=0.6]{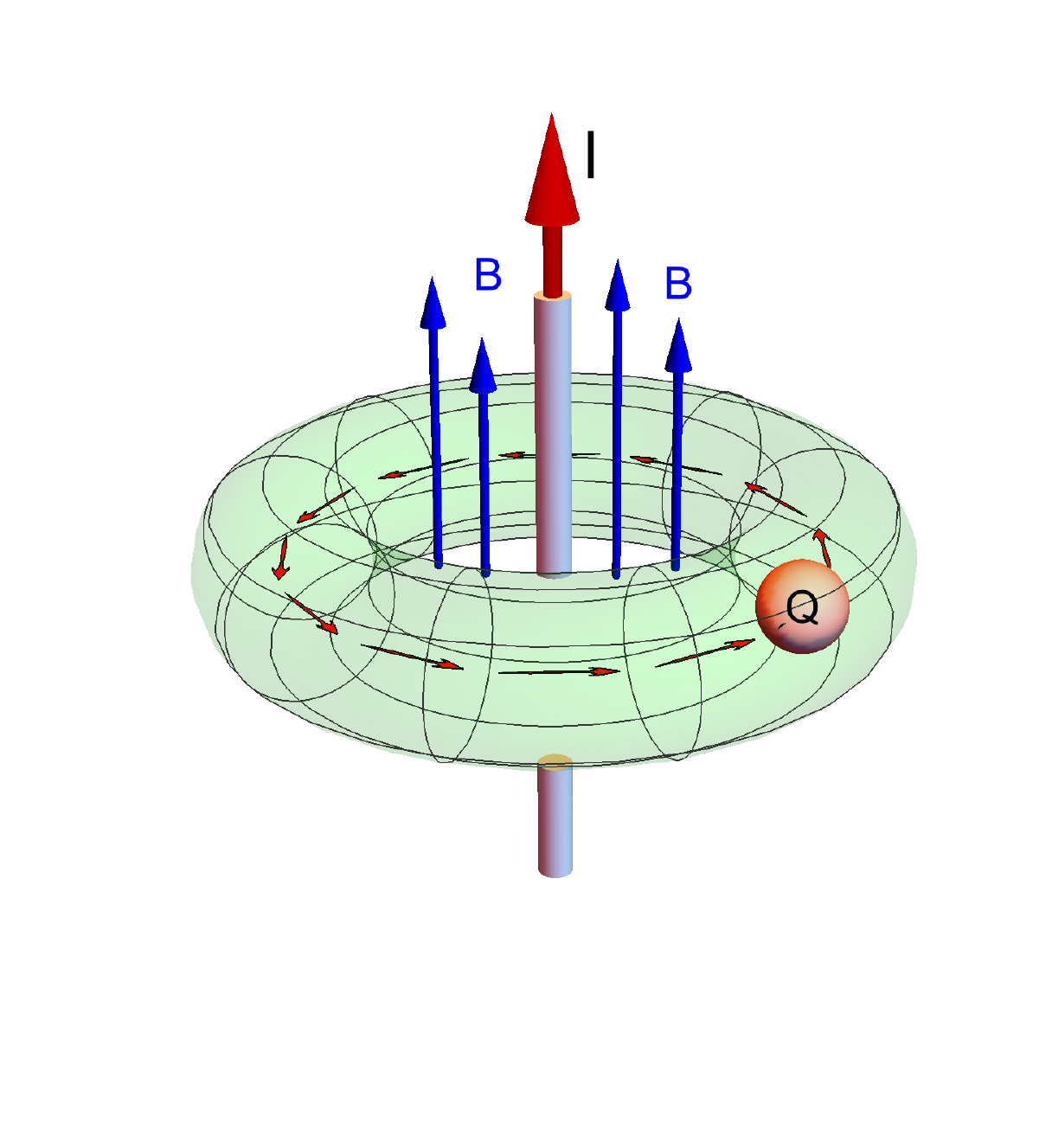}
\caption{\label{fig:fig10} Illustration of a toroidal trap in which an ion of charge $Q$ simulates the motion of a planar, quasi-rigid, rotor. A current $I$ (red arrow) threading the doughnut hole  induces an axial magnetic field. The system is
subjected to a background magnetic bias field (blue) arrow.}
\end{figure}
We thread an electric current along the symmetry axis piercing the doughnut hole to induce a magnetic field along the axial direction of the torus. Alternatively, an axial magnetic field can also be generated by joining a solenoid at its ends to form a torus (i.e. a micro-tokamak). In addition to the toroidal axial field,
generated by current $I$, a constant homogeneous bias magnetic field of magnitude $B_{0}$ parallel the symmetry axis is applied. The Hamiltonian for this system is
\beq
H = \frac{-\hbar^{2}}{2 m} ({\bm \nabla} \mathbb{1}  - i \frac{q}{\hbar c} {\bm A}_{0}  )^{2} + {\bm U}(\phi)  {\bm \Delta}(\rho) {\bm U}^{\dag}(\phi) + V_{trap} \nonumber \\
\eeq
where, in a cylindrical coordinate system, 
\beq
{\bm A}_{0} =  
\Bigl ({\hat {\bm \phi}} \,  \frac{B_{0} \, \rho}{2} + {\hat {\bm z} } \,
\frac{\mu_{0} I}{2 \pi} \ln\rho \Bigr ) \mathbb{1}
\label{U.02}
\eeq
is the Landau gauge vector potential for the total magnetic field.
$ {\bm U}(\phi)$ is given by Eq. (\ref{3.03}), 
$$ \cos\theta  =\frac{B_{0}}{ \sqrt{ {B_{0}}^2 + 
(\frac{ u_{0} I}{2 \pi \rho})^2 }}, $$
$q$ is the charge
of the ion, $\mu_{0}$ the magnetic constant, 
$$ {\bm \Delta}(\rho) = \mu \sqrt{{ B_{0}}^2 + (\frac{I \mu_{0}}{2 \pi \rho})^2 }  \, \sigma_{3}, $$
and $ V_{trap} $ is a trapping potential. 

In the adiabatic representation, and assuming that $V_{trap}$ is independent
of spin, we obtain the eigenvalue Schr{\"o}dinger equation,
\beq
&& \frac{-\hbar^{2}}{2 m} ({\bm \nabla} - i {\cal A})^{2} F({\bf r}) + 
  {\bm \Delta}(\rho) F({\bf r}) + \nonumber \\ 
&& V_{trap}({\bf r}) F({\bf r}) = E F({\bf r}) 
\label{U0.2}
\eeq
where
\beq
&& {\cal A} = i {\bm U}^{\dag} {\bm \nabla} {\bm U} + {\bm A}_{0} = \nonumber \\
&& \, \nonumber \\
&& \frac{ {\hat {\bm \phi} } }{2 \rho} \left ( \begin{array}{cc}  \cos\theta -1 + \frac{q}{\hbar c} B_{0} \rho^2 
& i \, \sin\theta \exp(-i \phi) \\
-i \, \sin\theta \exp(i \phi) & (1- \cos\theta) + \frac{q}{\hbar c} B_{0} \rho^2
\end{array} \right ) + \nonumber \\
&& {\hat {\bm \rho}} 
\left ( \begin{array}{cc}
     0 & -\frac{1}{2} \exp(-i \phi) \theta'(\rho) \\
     -\frac{1}{2} \exp(i \phi) \theta'(\rho) & 0 
\end{array} \right ) \nonumber \\
&& -{\hat {\bm z}} \,  \frac{q \,\mu_{0} I}{2 \, \hbar \, c \, \pi} \ln\rho 
\left (\begin{array}{cc}
    1 & 0 \\
    0 & 1
\end{array} \right ).
\label{U.04}
\eeq
Assuming that the trap potential is effective in freezing the degrees of freedom in the radial and ${\hat {\bm z}}$ direction, and for a large Zeeman energy gap ${\Delta}$, we replace the 3D Schr{\"o}dinger Eq. (\ref{U0.2})
with an effective 1D equation corresponding to a rigid planar rotor,
\beq
&& \frac{-\hbar^2}{ 2 m \rho_{0}^2}
(\partial_{\phi} - i {\cal A}_{eff})^2 F(\phi) 
+ {\bm \Delta}(\rho_{0}) F(\phi) = E \, F(\phi) \nonumber \\
&& {\cal A}_{eff} = \nonumber \\
&& \frac{1}{2} \left ( \begin{array}{cc}
     \cos\theta(\rho_{0}) -1 + \frac{q \Phi }{\hbar \, \pi \, c}   &  0\\
    0  &  1-\cos\theta(\rho_{0}) + \frac{q \Phi}{\hbar \, \pi \, c}
\end{array} \right ) \nonumber \\
\label{U.05}
\eeq
where $ \rho_{0} $ is the equilibrium value of the radial coordinate,
and $ \Phi = B_{0} \pi {\rho_{0}}^2 $ is the total magnetic flux
enclosed by the rotor. By tuning the current $I$ and the bias field
$ B_{0}$ we can alter and discriminate the values of the Wilson loop for different spin states. For example, if 
$$ \cos\theta(\rho_{0}) -1 +
\frac{q \Phi}{\hbar \pi c} =0 $$ then,
\beq
{\cal A}_{eff} \rightarrow \left ( \begin{array}{cc}
    0 & 0  \\
    0 & \frac{q \Phi}{\hbar \, \pi \, c} 
\end{array} \right ).
\label{U.06}
\eeq
In this scenario  the upper Zeeman level undergoes the motion of a free rotor, whereas 
the lower component experience an effective AB flux tube with charge
$ \Phi $. Such a capability, if realized, could find application as a novel magnetometer and rotational sensor.

The planar rotor has also been used as a model for the anyon\cite{Wilczek82}. In adiabatic transport about
a flux tube it can acquire a non-integer phase
(modulus $2 \pi$) as it completes one circuit. In the rotor systems discussed here
adiabatic transport is problematic as an initial wave packet
spreads in time. However, as a closed system, it eventually revives to its original shape.
For example, the
propagator for a spin-1/2 planar rotor coupled to a Wu-Yang flux tube of ``charge'' $\alpha$  is given by
\beq
&& G(\phi t; \phi' t'=0) = \nonumber \\
&& \sum_{m} \frac{1}{2 \pi} \exp(i m (\phi-\phi')) 
\exp(-i \frac{\hbar^2}{2 I} (m- \alpha \, \sigma_{3} )^2 t) 
\nonumber
\eeq
Or,
\beq
&& \exp(-i \frac{\hbar^2 \alpha^2 t}{2 I} ) \sum_{m} 
\frac{1}{2\pi} \exp(i m (\phi-\phi'+ \sigma_{3} \, 
\frac{\hbar^2 \, t \, \alpha}{I})) \times \nonumber \\
&& \exp(-i \frac{\hbar^2 \,m^2 \, t}{2 I} ).
\eeq
Now at the revival\cite{ROBINETT} time $t_{N} = 
\frac{4 \pi I \, N}{\hbar^2}  $, where $N$ is an integer,
\beq
&& G(\phi \, t_{N};\phi') = \nonumber \\ 
&&\frac{1}{2 \pi} \exp(i \Delta \phi_{N} \, \frac{\alpha}{2} ) \sum_{m} \exp(i m (\phi-\phi' + \sigma_{3} \Delta \phi_{N})) = \nonumber \\
&&  
\exp(i \Delta \phi_{N} \, \frac{\alpha}{2}) 
\left ( \begin{array}{cc}
     \delta(\phi-\phi'+\Delta \phi_{N})& 0 \\
     0 & \delta(\phi-\phi'-\Delta \phi_{N})
\end{array} \right ) \nonumber \\
\label{U0.7}
\eeq
where 
$ \Delta \phi_{N} = t_{N} \, \frac{\hbar^2 \alpha}{I} = 4 \pi N \alpha $. Thus an arbitrary initial, localized, wave packet is displaced, depending on its spin state,
by an amount $\pm \Delta \phi_{N}$. Suppose $ \alpha = m/p $ is a rational number where $p$ is even, then the packet returns to its original starting
point, i.e. $ \Delta \phi = 0 \, {\rm Mod} \, 2 \pi $ at $t_{N^*}$ for $ N^*=p/2 $. So if a localized packet at $t=0$ has the form
\beq
\psi(\phi,t=0) = \left ( \begin{array}{c}
     \psi_{u}(\phi)  \\
     \psi_{d}(\phi)
\end{array} \right ),
\label{U0.8}
\eeq
it evolves to
\beq
\psi(\phi,t_{N^*}) = \left ( \begin{array}{l}
     \exp(i W_{m}(\alpha) ) \psi_{u}(\phi)  \\
     \exp(-i W_{m}(\alpha) )\psi_{d}(\phi)
\end{array} \right )
\label{U0.9}
\eeq
where
\beq
&& W_{m}(\alpha) \equiv \oint_{m} d{\bm R} \cdot {\bm A}_{AB} \nonumber \\
&& {\bm A}_{AB} = {\hat {\bm\phi}} \frac{ \alpha}{R}
\label{U0.9a}
\eeq
is the argument of a Wilson loop integral with winding number
$ m $. A similar argument can be used when $p$ is odd. Expression (\ref{U0.9}) demonstrates that an arbitrary wave packet revives, up to a topological phase factor $\exp(i W_{m} \, \alpha \, \sigma_{3} )$, at its
initial position.

On a final note, at the time of writing I have become aware of recent literature in which similar themes, presented in this paper, are discussed. Synthetic gauge structures on a ring lattice
have been explored in \cite{Das}, and non-Abelian Wu-Yang structures have been observed in optical systems\cite{Yang1021,Chen19}

\begin{acknowledgments}
I wish to acknowledge support by the National Supercomputing Institute for use of the Intel Cherry-Creek computing cluster. Part of this work
was also made possible by support from a NSF-QLCI-CG grant 1936848.
\end{acknowledgments}

\appendix

\section{}

According to Eqs. (\ref{0.01a}) and (\ref{0.02}) the Schr{\"o}dinger equation for a rotor with unit radius
is
\beq
- \frac{1}{2 I} (\frac{\partial}{\partial \phi} - i {\bm a})^2 F + {\bm \Delta} \, F = i \frac{\partial F}{\partial t} 
\label{A0.03}
\eeq
where the gauge potential
\beq
{\bm a} = q \, \left ( \begin{array}{cc} \cos\theta  & \exp(i \gamma) \sin\theta  \\
 \exp(-i \gamma) \sin\theta  & -\cos\theta \end{array} \right ), 
\label{A0.02}
\eeq
where $q$ is an integer and $\theta, \gamma$ are parameters.
To solve for its energy spectrum
we let $ F =\frac{ \exp(i m \phi)}{\sqrt{2 \pi} } \, {\bm c} $ so that
\beq
\frac{ (m \mathbb{1} - {\bm a})^2}{2 \, I}  \, {\bm c} + \Delta \, {\bm c} = i {\dot {\bm c}} 
\label{A0.04}
\eeq
or  $ {\bm h}  \, {\bm c} = i \, {\dot {\bm c}} $
\beq
{\bm h} = \mathbb{1}\frac{(m^2 + q^2)  }{2 I}  -
 \frac{m}{I} \left ( \begin{array}{cc} \cos\theta - \frac{I}{m} \Delta & \exp(i \gamma) \sin\theta  \\
\exp(-i \gamma) \sin\theta & - \cos\theta + \frac{I}{m} \Delta   \end{array} \right ) \nonumber \\
\label{A0.05}
\eeq
where we used the fact that $ {\bm a}\cdot {\bm a} = q^2 \,\mathbb{1} $. 
The eigenvalues of ${\bm h}$ are 
\beq 
&& e(m) = e_{0}(m) \pm e_{1}(m) \nonumber \\
&& e_{0}(m) = \frac{m^2+q^2}{2 I} \nonumber \\
&& e_{1} = \, 
\frac{\sqrt{m^2 \, q^2 + I^2 \, \Delta^2 -2 I \, m \, q \, \Delta \cos\theta }}
{2 \, I}
\label{A0.07}
\eeq
and, the partition function,
\beq
{\cal Z} = 2 \, \sum_{m} \exp(-\beta \, e_{0}(m)) 
\cosh(\beta \, e_{1}(m) )
\label{A0.22}
\eeq
where $\beta$ is the inverse temperature.
Consider the limit $ \Delta << 1 $, in which 
\beq
{\cal Z} \rightarrow  2 \, \sum_{m} \exp(- \beta \, \frac{m^2+1}{2 \, I} ) \times \nonumber \\ 
\cosh (\beta (\frac{m}{I} - \Delta \cos\theta )).
\label{A0.23}
\eeq
Taking the Poisson transform of the r.h.s of Eq. (\ref{A0.23}), we find
\beq
{\cal Z} \rightarrow  2 \,\sqrt{\frac{2 \,\pi \,I}{\beta} } \sum_{m} \exp(-\frac{2\, I \, \pi^2 \, m^2}{\beta} ) \, \cosh(\beta \, \Delta \, \cos\theta). \nonumber \\
\label{A0.24}
\eeq
Thus, in this limit the partition function assumes the form
of a free rotor in the presence of a constant ``scalar'' potential $ \Delta \cos\theta $.

In the other extreme, $ I \, \Delta >> 1 $,  
\beq
&& {\cal Z} \rightarrow  2 \, \sum_{m} \exp(- \beta \, \frac{m^2+1}{2 \, I} ) \times \nonumber \\ 
&& \cosh (\beta (\Delta - \frac{m}{I} \cos\theta  ))
\label{A0.26}
\eeq
or, applying the Poisson summation formula,
\beq
&& {\cal Z} \rightarrow  2 \,\sqrt{\frac{2 \,\pi \,I}{\beta} }
\cosh(\beta \, \Delta) \exp(-\beta \frac{\sin^2\theta }{2 \, I}) 
\times \nonumber \\
&& \sum_{m} \exp(-\frac{2 \,\pi^2 \, m^2 \, I}{\beta}) \cos(2 \, \pi \, m \, \cos\theta).
\label{A0.27}
\eeq
In Fig. (\ref{fig:Appfig1}) we plotted
the logarithm of the ratio $ {\cal Z}/{\cal Z}_{0} $
where
$$ {\cal Z}_{0} \equiv 
2 \,\sqrt{\frac{2 \,\pi \,I}{\beta} }
\cosh(\beta \, \Delta) \exp(-\beta \frac{\sin^2\theta }{2 \, I}). $$
\begin{figure}[ht]
\centering
\includegraphics[scale=0.6]{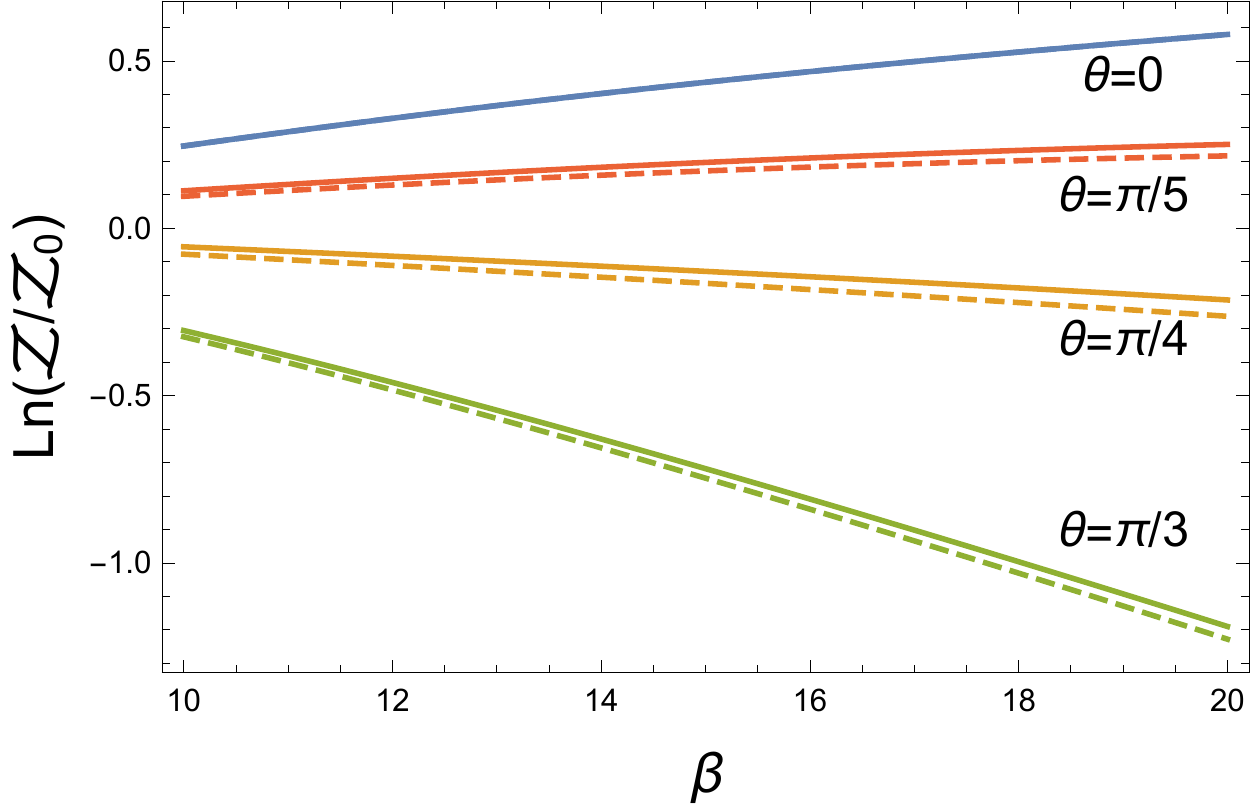}
\caption{\label{fig:Appfig1} Plot of ratio $Ln({\cal Z}/{\cal Z}_{0})$ as a function of the inverse temperature $\beta$. The values $I=1$, $\Delta=100$, were used to obtain this data.}
\end{figure}
In that figure the solid lines are calculated using
the exact values Eq. (\ref{A0.22}) for ${\cal Z}$, 
whereas the dashed lines represent the value
obtained using the approximate expression (\ref{A0.27}). According to Eq. (\ref{A0.27}), the ratio
\beq
{\cal Z}/{\cal Z}_{0} = 
 \sum_{m} \exp(-\frac{2 \,\pi^2 \, m^2 \, I}{\beta}) \cos(2 \, \pi \, m \, \cos\theta)
\nonumber
\eeq
in the limit $ \beta >> 1.$
The variation of this ratio, shown in Fig. (\ref{fig:Appfig1}), demonstrates the role
of the topological contribution $ \cos(2\, \pi \, m\, \cos\theta) $ to the, gauge invariant, partition function.

\section{}

According to Eq. (\ref{Z11}) the reduced partition function
\beq
&& z =  2  \sum_{m}  \exp(-\beta \, e_{0}) \cosh(\beta \,e_{1}) \nonumber \\
&& e_{0} = \frac{m^2+ \cos^2\theta}{2I}  \nonumber \\
&& e_{1} = \sqrt{ \Delta^2 +\frac{m^2}{I^2}\cos^2\theta }.
\label{A00.2}
\eeq
At cold temperatures as, i.e. $ \beta \rightarrow \infty $,
the approximation
\beq
&&  e_{1} \approx \alpha_{0} + \alpha_{1} | m|  \nonumber \\
&& \alpha_{0} = \Delta \nonumber \\
&& \alpha_{1} = -\Delta +  \sqrt{ \Delta^2 +\frac{\cos^2\theta}{I^2} }  
\label{A00.3}
\eeq
is appropriate.
Therefore, we need  to evaluate 
\beq
&& z = 2 \sum_{m} \exp(- \beta ( \frac{m^2 + \cos^2\theta}{2 I}) ) \times \nonumber \\
&& \cosh(\beta ( \alpha_{0} + | m | \alpha_{1} ) )
\label{A0.0}
\eeq
or
\beq
&& z =2 \exp(-\beta \frac{\cos^2\theta}{2 I} ) \sum \, \exp(-\beta \frac{m^2}{2 I}) \times \nonumber \\
&& \Bigl ( \cosh(\beta\, \alpha_{0} ) \cosh( \beta \, \alpha_{1} |m | ) +  \sinh(\beta \, \alpha_{0} ) \sinh( \beta \alpha_{1} |m | ) \Bigr ). \nonumber \\
\label{A0.1}
\eeq
The Poisson transform of Eq. (\ref{A0.1})  leads to
\beq
&&z= 2 \exp(-\frac{\beta \, \cos^2\theta}{2 I})\sqrt{\frac{2 \pi \, I}{\beta}} \Bigl (
\cosh(\alpha_{0}\beta )  \, \times \nonumber \\
&&   \sum_{k}   \exp(-\frac{2 \, \pi^2 \, I \, k^2}{\beta})
\exp(\frac{\alpha^{2}_{1} \, I \, \beta}{2}) \cos(2 \, \pi \, I \, k \, \alpha_{1}) - \nonumber \\
&& \frac{2}{\sqrt{\pi} } \sinh(\alpha_{0}\,  \beta ) \, \sum_{k} \,  {\rm Im} \, D_{F}(\sqrt{\frac{I}{2 \beta}} (2 \pi k  - i \, \alpha_{1} \, \beta) 
\Bigr )
\label{A0.2}
\eeq
where $D_{F}$ is the Dawson integral\cite{Dawson} and where the summation is over all integers $k$.
It is useful to express the latter in terms of a confluent hypergeometric function \cite{Dawson}
\beq
D_{F}(\xi) = \xi \, \exp(-\xi^2) \,  _{1}F_{1}(\frac{1}{2},\frac{3}{2} ,\xi^{2}). 
\label{A0a.1}
\eeq
For $ |\xi| >> 1 $ we use the asymptotic expansion for the Kummer function\cite{Stegun} 
\beq
_{1}F_{1}(\frac{1}{2},\frac{3}{2} ,\xi^{2})  \rightarrow \frac{\exp(\xi^2)}{2 \xi^2} \pm  \frac{i \sqrt{\pi}}{2 \sqrt{\xi^2}}  
\label{A0a.2}
\eeq
where the $ \pm $ sign refers to the cases
$$ -\frac{\pi}{2} <  arg(\xi^2) < \frac{3 \pi}{2} \quad \& \,   -\frac{3 \pi}{2} <   arg(\xi^2) \leq   -\frac{\pi}{2} $$
respectively. Or
\beq
D_{F}(\xi) \rightarrow \frac{1}{2 \, \xi } 
\pm \frac{i \sqrt{\pi} }{2} \, \exp(-\xi^2)
\label{A0aa.2}
\eeq
where $\pm$ corresponds to ${\rm Re}(\xi) > -{\rm Im}(\xi)$
and $ {\rm Re}(\xi) < - {\rm Im}(\xi) $ respectively.
Since $ \xi =  \sqrt{\frac{I}{2 \beta} }\, ( 2\pi \,k - i \, \beta \, \alpha_{1} ) $ we find that 
as $ \beta \rightarrow \infty $ ($\alpha_{1} \neq 0 $ )
\beq
&& {\rm Im} \, D_{F}\Bigl (\sqrt{\frac{I}{2 \beta}} (2 \pi k  - i \, \alpha_{1} \, \beta) \Bigr ) \rightarrow \nonumber \\
&& \sqrt{ \frac{\beta}{2 \, I }} \,
\frac{ \beta \, \alpha_{1} }{4 \pi^2 k^2+ \alpha_{1}^2 \beta^2 } \pm 
\nonumber \\
&& \frac{\sqrt{\pi}}{2}\exp(\frac{\alpha_{1}^2 \, I \,
\beta}{2} ) \exp(-\frac{2 \pi^2 \, I \, k^2}{\beta} ) \cos(2 \pi\, I \, k \, \alpha_{1} )
\label{A0a.3}
\eeq
Thus, if $\alpha_{0} > 0 $,
\beq
&& z \approx \sqrt{\frac{2 \pi \, I}{\beta}} \exp(-\frac{\beta \, \cos^2\theta}{2 I} ) \exp(\alpha_{0}\, \beta)\exp(\frac{\alpha^{2}_{1} \, I \, \beta}{2}) \times \nonumber \\
&&  
\sum_{k} \exp(-\frac{2 \pi^2 I \, k^2}{\beta} )
\cos(2 \pi \, I \, \alpha_{1} \, k) \, g(k)  
\label{A0.3}
\eeq
where 
\beq 
&& g(k) = 2 \quad {\rm for} \quad 2\pi \, k < \alpha_{1} \, \beta \nonumber \\
&& g(k) = 0 \quad {\rm for} \quad 2\pi \, k > \alpha_{1} \, \beta
\nonumber
\eeq
and we used the fact
\beq
\sum_{k} \frac{\alpha_{1} \beta}{ 4 \, \pi^2 \, k^2 +\alpha_{1}^2 \beta^{2} }=\frac{1}{2}
\coth\frac{\alpha_{1}\, \beta}{2} \approx \frac{1}{2} 
\label{A0.4}
\eeq
in this limit. Using definitions (\ref{A00.3}) 
 and so
$$ \alpha_{0}=\frac{\sin^2\theta}{4 \, I} \quad \quad 
\alpha_{1} = -\frac{\sin^2\theta}{4 \, I} + \frac{\sqrt{\frac{\sin^2\theta}{16} + 
\cos^2\theta }}{I} , $$
and $ \Delta = \frac{\sin^2\theta}{4 \, I} $ we find that,
\beq
&& z \rightarrow  2\sqrt{\frac{2 \, \pi I}{\beta}} \,\exp(\beta \, V(\theta) ) \times \nonumber \\
&& \sum_{k}   \exp(-S_{0}(k)) 
\cos(2 \, \pi \, k \, \Omega )
\label{A0.5}
\eeq
where $ S_{0}(k)$ is the Wick rotated action for a free rotor undergoing $k $ circuits
and
\beq
&& V(\theta) = -\frac{\cos^2\theta}{I} + \alpha_{0} + \frac{ {I \, \alpha_{1}}^2}{2} \nonumber \\
&& \Omega =  -\frac{\sin^2\theta}{4} + \sqrt{\frac{\sin^2\theta}{16} + \cos^2\theta }.
\label{A0.6}
\eeq

\section{}
We first demonstrate that a particle in the presence of a quantized gauge field begets a multicomponent
wave equation whose amplitudes are coupled to a non-Abelian gauge potential.
As an example, consider the Hamiltonian for a charged (first quantized) particle coupled to a quantized, transverse, Maxwell gauge field,
\beq
&& H = \frac{1}{2 \, m} ({\bm p} - {\bm A}_{r} )^2 + \sum_{k\, \lambda} \, \hbar \omega_{k \lambda } \, a_{{\bm k} \lambda}^{\dagger} a_{{\bm k} \lambda } 
  \nonumber \\
&& {\bm A}_{r} = \sum_{k \, \lambda} \Bigl ( {{\bm A}^*}_{{\bm k}\lambda} \, a_{{\bm k}\Lambda}  +  {{\bm A}^*}_{{\bm k}\lambda} \, {a}^{\dagger}_{k\Lambda} \Bigr )
\label{c00.0}
\eeq
Here $ {\bm p}$ is the particle momentum operator conjugate to ${\bm r}$. $ a_{{\bm k}\lambda}, a_{{\bm k}k\lambda}^{\dagger} $ are, respectively, photon
destruction and creation operators that satisfy commutation relations $ [a_{{\bm k} \lambda}, {a^{\dagger}}_{{\bm k}\lambda}] =
\delta_{{\bm k},{\bm k}'} \delta_{\lambda,\lambda'}, $  and  ${{\bm A}^*}_{{\bm k}\lambda} $ is an amplitude  for a photon with momentum ${\bf k}$, and
polarization ${\lambda}$. For the sake of simplicity, and without loss of generality, we consider only single mode field quanta that are eigenstates
of the number operator $ a^{\dagger} a$, where we supressed the mode index. 
The eigenstates of the radiation field 
$$H_{rad} \equiv \hbar \omega \, a^{\dag} \,  a \,  $$
are labeled by the occupation number and so an eigenstate of Hamiltonian (\ref{c00.0}), can always be
written as a linear combination
\beq
\Psi = \sum_{n} \, f_{n}({\bm r}) | n \rangle  
\label{c00.1}
\eeq
where $| n \rangle $ is an eigenstate of the number operator $  a^{\dag} \,  a \ $  and $n$ is
the occupation number.  Using expression (\ref{c00.1}) and treating the amplitudes
$f_{n}({\bm r})$ as variational parameters we arrive, using the fact that the set
$| n \rangle $ are orthonormal, the set of coupled equations,
\beq
\frac{1}{2 m}( \nabla - i \, {\underline {\bm A}}_{r} )^2 \, {\underline F}({\bf r}) + 
{\underline V} \, {\underline F}({\bf r})  = E \,  {\underline F}({\bf r}).
\label{c00.2}
\eeq
Here 
\beq
{\underline F}({\bf r}) \equiv \left( \begin{array}{c} f_{1}({\bf r}) \\
    f_{2}({\bf r}) \\
     \vdots \\
 \end{array} \right )
\label{c00.3}
\eeq
is an infinite dimensional column matrix. 
$ {\underline {\bm A}_{r}} $  is  a square matrix  whose $n \, m$th entry
$ {\bm A}_{nm}  = \langle n | {\bm A}_{r} | m \rangle $, and $ {\underline V} $ is a diagonal
matrix whose $n$th  entry is $ n \, \hbar \omega$.

Consider a Hilbert space generated by 
bosonic operators ${ a}, { a}^{\dag} $ so that  $ [a, a^{\dagger}] = 1 $
This space is spanned by the basis vectors
\beq 
| n \rangle = \frac{ (a^{\dag})^n}{\sqrt{n!}} \, | 0 \rangle 
\label{d00.1}
\eeq
where $a | 0 \rangle =0$.
In this space we define a Hamiltonian 
\beq
{  H}_{BO} = e({  a}^{\dag} \, {a}) 
\label{d00.3}
\eeq
where $e$ is an arbitrary function. The spectrum  of $H_{BO}$ is $e(n) $ for $ n \in  \mathbb{Z} \ge 0 $.

We now posit the Hamiltonian 
\beq
{  H} = \frac{ {\bm p}^2}{2 \, m} + {  U} \, {  H}_{BO} {  U}^{\dagger},  
\label{d00.2}
\eeq
which a straight-forward generalization of the finite dimensional 
models discussed in the main section. Here,
$ U$ is a unitary operator that, in general, is a function of ${\bm r}$ and ${ a}, {a}^{\dag}$.

For example, let
\beq
&& { U} =  \exp(-i \phi \, {  a}^{\dag} \, {  a})  \exp(-i \lambda ({  a} + {  a}^{\dag})  )
\exp(i \phi \, {  a}^{\dag} \, {  a}),
\label{d00.3a} 
\eeq
where $\phi$ is the azimuthal angle in a cylindrical coordinates system, and $\lambda$ is a real valued parameter.
Because the eigenvalues of the number operator ${  a}^{\dag} \, {  a} $ are integers, ${  U}$ is single valued, i.e.
$ {  U}(\phi=0) = {  U}(\phi=2 \pi) $, and so we can express the system amplitude
\beq
\Psi = \sum_{n} f_{n}({\bm r}) {\bm U} | n \rangle. 
\label{d00.4}
\eeq
 Using this ansatz we arrive
at the set of equations (\ref{c00.2}) where now the amplitudes $f_{n}({\bf r}) $ are coupled to
\beq
&& {\underline V}_{nm}  = \langle n | H_{BO}  | m \rangle  = e(n) \delta_{n\,m} \nonumber \\
&& {\underline {\bm A}}_{nm} = \langle n | {\bm A} | m \rangle \nonumber \\
&& {\bf A} =\frac{ i  {\hat {\bm \phi}}  }{r}  \, U^{\dag}  \partial_{\phi} U = 
 \frac{i {\hat {\bm \phi}}}{r} \Bigl ( -i \, U^{\dag} \, a^{\dag} a \, U + i \,  a^{\dag} a  \Bigr ) = \nonumber \\
&&  {\hat {\bm \phi}} \, \frac{i \lambda }{r} \Bigl (a \, \exp(i \phi) -a^{\dag} \, \exp(-i \phi) \Bigl ) +  {\hat {\bm \phi}} \, \frac{ \lambda^2 }{r} .
\label{d00.6}
\eeq
Here $ {\bm A}$ describes a pure gauge. Alternatively,
we could induce a unitary transformation
\beq
&& H' = U^{\dag} H U =  U^{\dag} \frac{ {\bm p}^2}{2 \, m} U +  {  H}_{BO} = \nonumber \\
&& \frac{1}{2 m}  ({\bm p} - \, {\bm A} )^2 + e(a^{\dag} a) 
\label{d00.7}
\eeq
so that $H'$ describes a particle minimally coupled to a dynamical Abelian gauge field ${\bm A}$.
In this picture the ansatz $  \sum_{n} f'_{n}({\bm r}) | n \rangle $ leads to identical equations for
the amplitudes $f'_{n}({\bm r})$ described above.

Allthough $ {\bm A}$ is a pure gauge, low energy eigensolutions to $H'$ exhibit, as we demonstrate below, non-trivial effective gauge structure.  For example, suppose that
$ e(n) >> e(0) $ for $n>0$. We can then employ the PSS approximation, which begets the Schr{\"o}dinger equation
\beq
&& \frac{1}{2 m} ({\bm \nabla} - i {\bf A}_{eff} )^{2}  F({\bf r}) +  {\underline V}_{eff}  F({\bf r}) = E \, F({\bf r}) \
\label{d00.7a}
\eeq
for the ground state scalar amplitude  $F({\bf r}) $. Here
\beq
&&  {\bm A}_{eff} = \hat {\bm \phi} \frac{ \lambda^2 }{r}  
\label{d00.8}
\eeq
\begin{figure}[ht]
\centering
\includegraphics[scale=0.3]{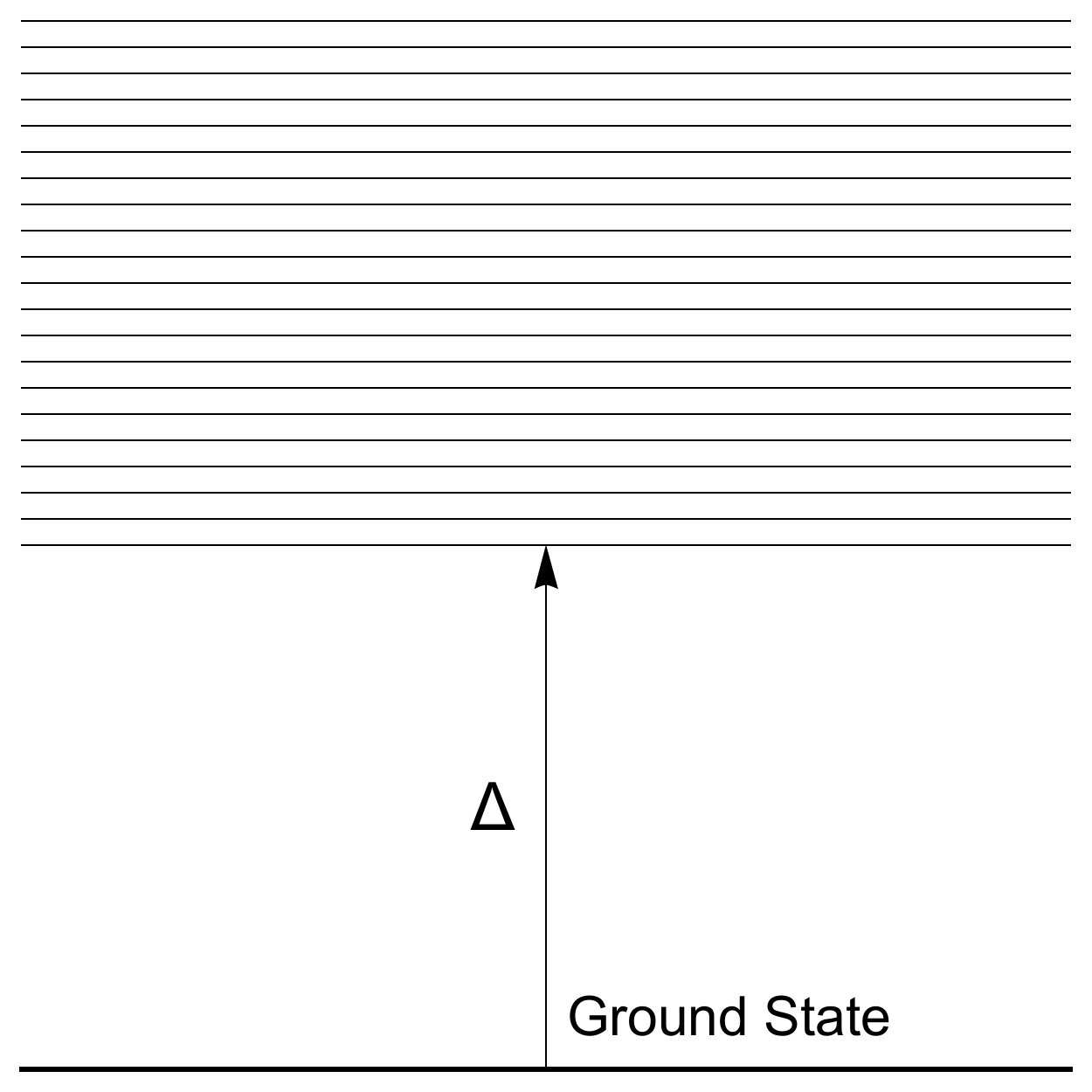}
\caption{\label{fig:appfig3} }
\end{figure}

 is the gauge potential of an Aharonov-Bohm flux tube of charge $\lambda^2$, 
and
\beq
  {\underline V}_{eff} = e_{0} + \frac{1}{2 m} \sum_{n \neq 0}{\bm A}_{0n} \cdot {\bm A}_{n0}.
\label{d00.9}
\eeq
is an effective scalar potential that is the sum of the adiabatic ground state energy 
 $e_{0} \equiv e(0) = \langle 0 | e(a^{\dag} a) |  0 \rangle $
and the correction
\beq
&& \frac{1}{2 m} \sum_{n \neq 0}{\bm A}_{0n} \cdot {\bm A}_{n0} = \nonumber \\
&& \frac{\lambda^2}{2 m r^2} \, \langle 0 | a \exp(i\phi) | 1 \rangle \langle  1 | \exp(-i\phi) a^{\dag} | 0 \rangle = \nonumber \\
&&  \frac{1}{2 m} \frac{\lambda^2}{r^2}.
\label{d00.10}
\eeq
We can think of the latter as a self-energy induced by the emission and re-adsorption of gauge quanta, thus demonstrating dynamical content encapsulated in ${\bm A}$.

\bibliographystyle{apsrev4-1}
\bibliography{bib15}

\begin{thebibliography}{46}
\bibitem [{com()}]{comment01}%
  \BibitemOpen
  \href@noop {} {}\bibinfo {note} {The Wilson loop integral for a WY flux tube
  is twice that of a single AB flux tube}\BibitemShut {NoStop}%
\end{thebibliography}

\end{document}